\PassOptionsToPackage{strict}{changepage} 
\documentclass[universe,review,accept,pdftex,moreauthors]{Definitions/mdpi} 
\usepackage[utf8]{inputenc} 
\usepackage{tabularx}   
\usepackage{ragged2e}   
\usepackage{caption}    
\usepackage{calc}       
\usepackage[strict]{changepage} 
\newcolumntype{L}{>{\RaggedRight\arraybackslash}X} 

\usepackage{longtable} 

\firstpage{1} 
\pubvolume{12}
\issuenum{1}
\articlenumber{24}
\pubyear{2026}
\copyrightyear{2026}
\externaleditor{Ioana Dutan} 
\datereceived{17 November 2025} 
\daterevised{28 December 2025 } 
\dateaccepted{8 January 2026 } 
\datepublished{15 January 2026  } 
\hreflink{https://doi.org/} 

\def\hl{}

\Title{{Future} 
Perspectives on Black Hole Jet Mechanisms: Insights from Next-Generation Observatories and Theoretical Developments}

\Author{\hl{Andre L. B. Ribeiro} 
 *\orcidA{} and \hl{Nathalia M. N. da Rocha} 
 *\orcidB{}}

\AuthorNames{Andre L. B. Ribeiro and Nathalia M. N. da Rocha}

\address[1]{Departmento de Ciências Exatas, Universidade Estadual de Santa Cruz, Rodovia Jorge Amado km 16, \mbox{Ilhéus 454650-000, Bahia, Brazil }}

\corres{\hangafter=1 \hangindent=1.05em \hspace{-0.82em}Correspondence: albr@uesc.br (A.L.B.R.); nmnrocha@uesc.br (N.M.N.d.R.)}

\abstract{Black hole jets represent one of the most extreme manifestations of astrophysical processes, linking accretion physics, relativistic magnetohydrodynamics, and large-scale feedback in galaxies and clusters. Despite decades of observational and theoretical work, the mechanisms governing jet launching, collimation, and energy dissipation remain open questions. In this article, we discuss how upcoming facilities such as the Event Horizon Telescope (EHT), the Cherenkov Telescope Array (CTA), the Vera C. Rubin Observatory (LSST), and the Whole Earth Blazar Telescope (WEBT) will provide unprecedented constraints on jet dynamics, variability, and multi-wavelength signatures. Furthermore, we highlight theoretical challenges, including the role of magnetically arrested disks (MADs), plasma microphysics, and general relativistic magnetohydrodynamic (GRMHD) simulations in shaping our understanding of jet formation. By combining high-resolution imaging, time-domain surveys, and advanced simulations, the next decade promises transformative progress in unveiling the physics of black hole jets.}


\begin{document}



\section{Introduction}

Relativistic jets launched by accreting black holes are among the most energetic and enigmatic phenomena in astrophysics. These collimated outflows, capable of propagating across intergalactic distances, are thought to originate in the immediate vicinity of {the event horizon} 
~\citep{10.1093/mnras/179.3.433, 10.1093/mnras/199.4.883, 10.1007/978-3-642-01936-4}. Yet, despite extensive observational campaigns and decades of theoretical modeling, the~physical mechanisms underlying jet launching, collimation, and~energy dissipation remain unsettled. Addressing these open questions is crucial not only for understanding black hole physics but also for unraveling the impact of active galactic nuclei (AGN) feedback on galaxy formation and evolution~\citep{10.1146/annurev-astro-081811-125521, 10.1146/annurev-astro-081913-035722}.

The study of black hole jets spans multiple scales and physical regimes, from~quantum mechanical processes near the event horizon to large-scale hydrodynamic interactions with the intergalactic medium~\citep{1999agnf.book.....K, 10.1111/j.1745-3933.2007.00413.x}. At~the smallest scales, the~ergosphere of a rotating black hole provides the energy reservoir that powers these outflows through the Blandford-Znajek mechanism~\citep{10.1093/mnras/179.3.433}. The~efficiency of this process depends critically on the magnetic field configuration threading the black hole, with~magnetically arrested disks (MADs) representing the most efficient regime for energy extraction~\citep{10.1088/0004-637X/711/1/50, 10.1111/j.1365-2966.2012.21074.x}. However, the~transition from magnetospheric processes to hydrodynamic jet propagation involves complex plasma physics that remains poorly understood~\citep{10.1111/j.1365-2966.2007.12050.x, 10.1088/0004-637X/698/2/1570}.

\textls[-27]{Observationally, jets manifest across the electromagnetic spectrum, from~radio wavelengths where synchrotron emission traces the bulk motion of relativistic particles to~gamma-ray energies where inverse Compton scattering and potentially hadronic processes dominate~\citep{10.1086/133630, 10.1016/j.crhy.2015.07.003}. The~multi-wavelength nature of jet emission yields complementary diagnostics of the underlying physics, but~also presents challenges in developing unified theoretical frameworks~\citep{10.1088/0004-637X/768/1/54}. For~instance, the~rapid variability observed in blazars at TeV energies implies emission regions smaller than the gravitational radius of the central black hole, challenging our understanding of particle acceleration and energy dissipation in relativistic outflows~\citep{10.3847/2041-8205/824/2/L20, 10.1086/520635}.}

Recent years have witnessed remarkable advances in jet studies across multiple fronts. The~Event Horizon Telescope (EHT) captured horizon-scale imaging of the supermassive black hole M87*, revealing jet-launching structures with unprecedented detail~\citep{10.3847/2041-8213/ab0ec7, 10.3847/2041-8213/ab0f43, 10.3847/2041-8213/abe71d}. These observations confirmed theoretical predictions about the existence of a bright ring structure surrounding the black hole shadow and established direct evidence for the launching of jets from the immediate vicinity of the event horizon~\citep{10.1093/mnras/staa922}. The~asymmetric brightness distribution observed in M87* suggests the presence of ordered magnetic fields and places constraints on the black hole spin and viewing angle~\citep{10.3847/2041-8213/ab1141, 10.3847/2041-8213/ab518c}.

\textls[-35]{High-energy observations from instruments such as Fermi-LAT  and H.E.S.S. (High Energy Stereoscopic System) have traced variability in blazar jets down to minute timescales, pointing toward compact emission regions and challenging classical shock-based acceleration models~\mbox{\citep{10.3847/2041-8205/824/2/L20, 10.1051/0004-6361/201629419, 10.1086/521382}}. These observations reveal that particle acceleration can occur on timescales much shorter than the light-crossing time of the broad-line region, suggesting that acceleration occurs in the jet rather than in external photon fields~\citep{10.1111/j.1365-2966.2009.15898.x, 10.1111/j.1745-3933.2010.00884.x}. The~detection of orphan flares—gamma-ray flares without corresponding optical counterparts—further supports models where high-energy emission originates in spine-sheath jet structures with different Doppler factors~\citep{10.1088/2041-8205/710/2/L126, 10.1086/444593}.}

At the same time, general relativistic magnetohydrodynamic (GRMHD) simulations have achieved new levels of sophistication, allowing the exploration of MADs and their efficiency in extracting rotational energy from spinning black holes~\citep{10.1088/0004-637X/711/1/50, 10.1093/mnras/stac285, 10.3847/1538-4357/ab0c0c}. These simulations have revealed that the jet power can exceed the accretion luminosity when the magnetic flux threading the black hole reaches saturation levels~\citep{10.1016/j.newast.2010.03.001, 10.1111/j.1365-2966.2012.21074.x}. The~MAD state is characterized by periodic magnetic flux eruptions that modulate the jet power and may explain the observed variability in AGN and X-ray binaries~\citep{10.1093/mnrasl/slx174, 10.3847/2041-8213/ac46a1}.

However, many key questions remain unanswered: What determines the efficiency of jet launching? Current models suggest that both black hole spin and magnetic field strength play important roles, but~the relative importance of these factors and their interplay with accretion rate remains unclear~\citep{10.1126/science.1227416, 10.1093/mnras/staa476}. How is energy dissipated along the jet? While magnetic reconnection, shocks, and~turbulence have all been proposed as dissipation mechanisms, their relative contributions and spatial distribution along the jet are not well-determined~\citep{10.1088/2041-8205/783/1/L21, 10.1088/0004-637X/809/1/38}. What role does plasma composition play? The electron-positron versus electron-proton composition of jets affects their radiative signatures and propagation properties, but~observational evidence remains limited~\citep{10.1038/26675, 10.1093/mnras/stab163}. How do jets couple to their environments? The interaction between jets and their surrounding medium determines the morphology of radio lobes and the efficiency of AGN feedback, but~the physics of this coupling involves complex multi-phase processes that are challenging to model~\citep{10.1016/j.newar.2020.101539, 10.1093/mnras/sty067}.

The upcoming decade is poised to bring transformative progress in addressing these questions. Several next-generation observatories are expected to open new observational windows into black hole jet physics. The~continued development of the EHT will allow multi-frequency and time-resolved imaging of jet-launching regions, probing magnetic field topologies and plasma dynamics at horizon scales~\citep{10.48550/arXiv.1909.01411, 10.1126/sciadv.aaz1310}. Future additions to the EHT array, including space-based telescopes, will extend the baseline coverage and make possible imaging of Sagittarius A*, allowing for comparative studies of jet launching in different environments~\citep{10.3847/1538-4357/ab86ac, 10.1051/0004-6361/201936622}.

The Cherenkov Telescope Array (CTA) will revolutionize high-energy astrophysics, enabling population studies of blazars and time-domain analyses of jet flares with unprecedented sensitivity~\citep{10.1142/10986, 10.1007/s10686-011-9247-0}. CTA's improved angular resolution and sensitivity will allow detailed studies of jet structure and particle acceleration processes. Its wide field of view and rapid slewing capabilities will be key for follow-up observations of transient events, potentially revealing new classes of jet-powered phenomena~\citep{10.1016/j.astropartphys.2013.01.007}.

Meanwhile, the~Vera C. Rubin Observatory's Legacy Survey of Space and Time (LSST) will deliver long-term high-cadence monitoring of AGN variability across the sky, offering statistical insights into jet duty cycles and feedback processes~\citep{10.48550/arXiv.0912.0201, 10.3847/1538-4357/ab042c}. LSST's unprecedented combination of depth, area, and~cadence will drive the discovery of rare transients and place place statistical limits on jet triggering mechanisms and their dependence on host galaxy properties~\citep{10.48550/arXiv.1811.06542,10.1088/0004-637X/753/2/106}.

The Whole Earth Blazar Telescope (WEBT) collaboration exemplifies the power of global, coordinated, multi-wavelength monitoring to probe the extreme and dynamic physics of relativistic jets. Two recent studies — Jorstad~et~al.~\citep{10.1038/s41586-022-05038-9} and Webb \& Sanz~\citep{10.3390/galaxies11060108} — illustrate the extraordinary 2020 WEBT campaign on BL Lacertae to unveil connections across temporal and spatial scales, from~hours-long quasi-periodic oscillations (QPOs) to minute-scale micro-variability. WEBT will provide a multi-scale picture where large-scale jet instabilities and small-scale turbulent plasma processes are intrinsically~linked.

Together, these facilities promise to bridge the gap between horizon-scale physics and kiloparsec-scale jet propagation. The~synergy between high-resolution imaging, time-domain surveys, and~multi-wavelength observations will produce a comprehensive view of jet physics across all relevant scales and timescales~\citep{10.1146/annurev-astro-081817-051948}.

Complementing these observational breakthroughs, theoretical models face the challenge of incorporating increasingly complex plasma physics into relativistic frameworks. While GRMHD simulations have been instrumental in revealing the role of large-scale magnetic fields, they often neglect kinetic processes such as particle acceleration and radiative feedback, which are essential for connecting simulations with observations~\citep{10.1093/mnras/stv2084, 10.3847/0004-637X/829/1/11}. The~assumption of ideal Magnetohydrodynamic (MHD) breaks down on scales comparable to the plasma skin depth or gyroradius, where kinetic effects become important for particle acceleration and magnetic reconnection~\citep{10.1103/PhysRevLett.106.195003, 10.1093/mnras/stx2530}.

Hybrid approaches that combine fluid dynamics with particle-in-cell (PIC) simulations are emerging as a promising pathway to bridge this gap~\citep{10.1088/0004-637X/771/1/54, 10.1093/mnras/sty2636, 10.1103/PhysRevLett.118.055103}. These methods allow the self-consistent treatment of particle acceleration while maintaining the computational efficiency needed to model large-scale jet propagation. Recent advances in computational techniques, including adaptive mesh refinement and GPU acceleration, are making such hybrid simulations increasingly feasible~\citep{10.3847/1538-4365/aab114, 10.1029/2018JA025713, 10.1007/s11214-025-01142-0}.

\textls[-35]{Furthermore, advances in numerical relativity and high-performance computing will allow simulations to explore parameter spaces relevant for diverse astrophysical systems, from~stellar-mass black holes in X-ray binaries to supermassive black holes in AGN~\citep{10.12942/lrr-2008-7, 10.1103/PhysRevD.82.084031}. The~development of new numerical schemes that can handle the extreme dynamic range required to model jets—from the event horizon to kiloparsec scales—remains an active area of~research~\citep{10.3847/0067-0049/225/2/22, 10.3847/1538-4365/ac9966}.}

This paper reviews the key opportunities and challenges in the study of black hole jets, focusing on how next-generation instruments and theoretical developments can together transform our understanding of jet physics. We emphasize the multi-scale nature of the problem and the need for interdisciplinary approaches that combine observational astronomy, plasma physics, and~computational modeling. Section~\ref{sec2} will outline the main observational perspectives from the EHT, CTA, LSST, and~WEBT. Section~\ref{sec3} will discuss jet diversity across different astrophysical systems. Section~\ref{sec4} will cover the current state and future prospects of theoretical models, with~an emphasis on GRMHD simulations, plasma microphysics, and~jet-environment interactions. Finally, Section~\ref{sec5} will present open questions and future directions in the~field.

\section{Observational~Perspectives}\label{sec2}

The study of black hole jets requires observations across multiple wavelengths, spatial and time scales. The~next generation of astronomical facilities promises to unlock unprecedented insights into jet physics through complementary observational approaches. This section discusses the key capabilities and scientific prospects of three transformative facilities: the Event Horizon Telescope, the~Cherenkov Telescope Array, the~Vera C. Rubin Observatory, and~the Whole Earth Blazar~Telescope.

\subsection{The Event Horizon Telescope (EHT)}

\begin{figure}[H]
\includegraphics[width=95mm]{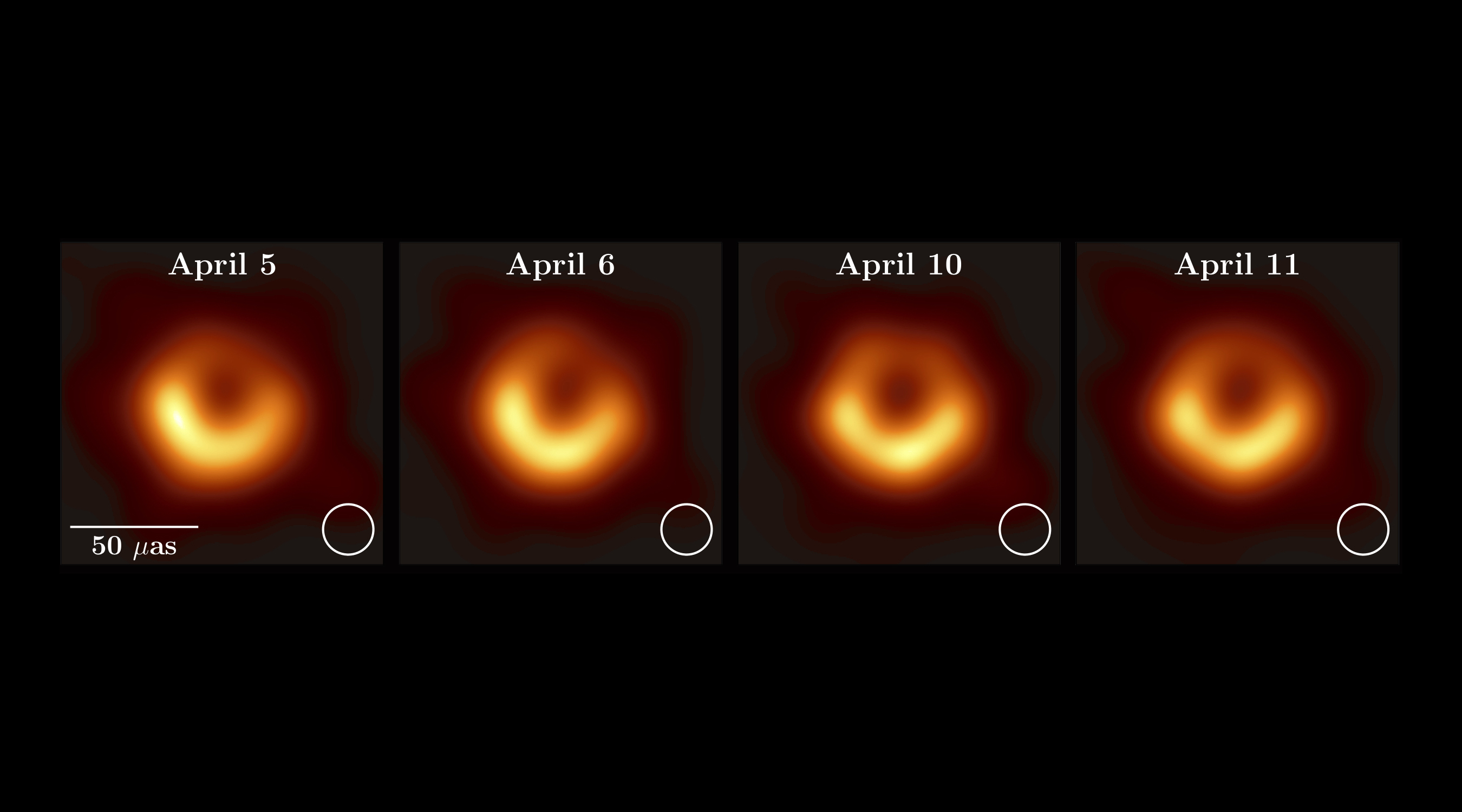}\\
\includegraphics[width=95mm]{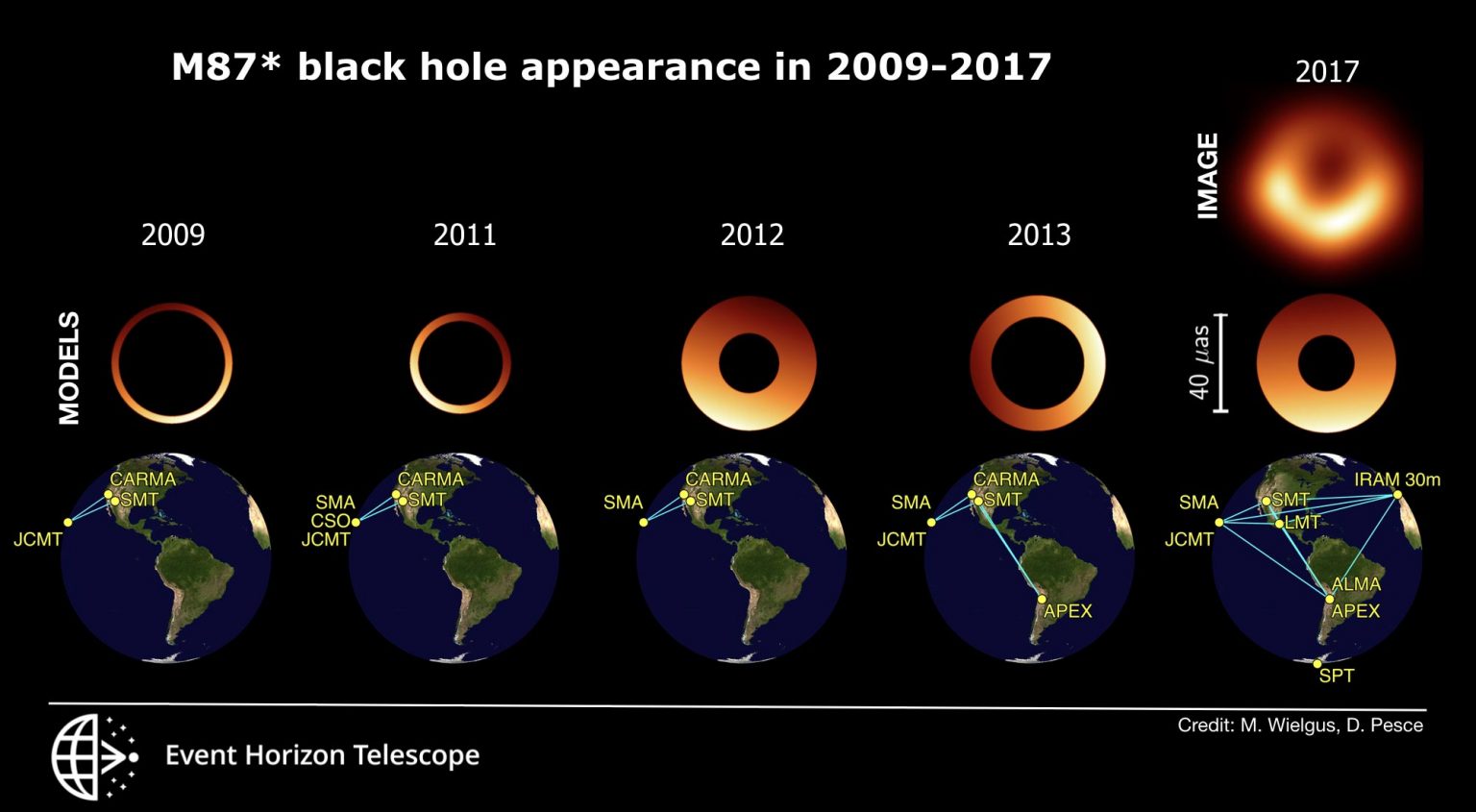}
\caption{\textls[-15]{(\textbf{Top}) \hl{Observations from} 
 the Event Horizon Telescope of the supermassive black hole at the center of
the elliptical galaxy M87, for~four different days. (\textbf{Bottom}) Snapshots of the M87* black hole appearance, obtained from
the EHT array of telescopes in 2009--2017. Where JCMT (James Clerk Maxwell Telescope), CARMA (Combined Array for Research in Millimeter-wave Astronomy), SMT (Heinrich Hertz Submillimeter Telescope), SMA (Submillimeter Array), CSO (Caltech Submillimeter Observatory), APEX (Atacama Pathfinder Experiment), LMT (Large Millimeter Telescope), IRAM (Institute for Radio Astronomy in the Millimetre Range) and SPT (South Pole Telescope). (EHT Collaboration (2019)~\citep{10.3847/2041-8213/ab0c96}).}}
\label{fig1}
\end{figure}

The EHT represents a milestone in high-angular-resolution astronomy, achieving micro-arcsecond imaging through very-long-baseline interferometry at millimeter wavelengths~\citep{10.1126/science.1224768, 10.3847/2041-8213/ab0ec7}. Its first results on M87* demonstrated the feasibility of resolving horizon-scale structures, including the shadow of the black hole and the jet-launching region~\citep{10.3847/2041-8213/ab0ec7, 10.3847/2041-8213/ab0f43}---see Figure~\ref{fig1}. These observations revealed the first direct visual evidence of a black hole and confirmed key predictions of general relativity in the strong-field regime~\citep{10.1103/PhysRevLett.125.141104, 10.1088/0004-637X/754/2/133}.

The success of the EHT M87* observations has opened new avenues for studying black hole physics and jet launching mechanisms. The~observed asymmetric ring structure imposes restrictions on the black hole spin, the~viewing angle, and~the magnetic field configuration in the immediate vicinity of the event horizon~\citep{10.3847/2041-8213/ab1141, 10.3847/2041-8213/ab518c}. The~brightness asymmetry is consistent with relativistic beaming effects expected from rotating plasma in the ergosphere, supporting models where jets are launched through the Blandford-Znajek mechanism~\citep{10.3847/1538-4357/abf117, 10.1093/mnras/stz2552}.

Future upgrades to the EHT will significantly enhance its scientific capabilities. Expanded frequency coverage, particularly observations at 345 GHz, will provide complementary information about the spectral properties of the emission region and help determine the magnetic field strength through synchrotron spectroscopy~\citep{10.48550/arXiv.1909.01411, 10.3847/1538-3881/abc3c3}. The~higher frequency observations are less affected by interstellar scattering, potentially revealing finer details of the jet launching region~\citep{10.1126/sciadv.aaz1310, 10.3847/1538-4357/aaf732}. Additionally, 345 GHz observations of Sagittarius A* may be feasible despite the strong scattering in the Galactic center, offering a second target for horizon-scale imaging~\citep{10.1086/373989, 10.3847/0004-637X/824/1/40}.

Increased baseline sensitivity through the addition of new telescopes and improved instrumentation will allow for time-resolved imaging of variability in jet bases~\citep{10.48550/arXiv.1909.01411, 10.1126/sciadv.aaz1310}. This capability is crucial for identifying magnetic flux structures, testing models of MADs, and~probing the coupling between black hole spin and jet power~\citep{10.1093/mnrasl/slx174, 10.3847/2041-8213/ac46a1}. The~characteristic timescales of variability near the event horizon are expected to be on the order of the gravitational timescale ($GM/c^3$), which corresponds to minutes for M87* and seconds for Sagittarius A*~\citep{10.1126/science.aav8137, 10.3847/1538-4357/abac0d}. Detecting and imaging such rapid variability will require coordinated observations with high temporal resolution and~sensitivity.

\begin{figure}[H]
\includegraphics[width=98mm]{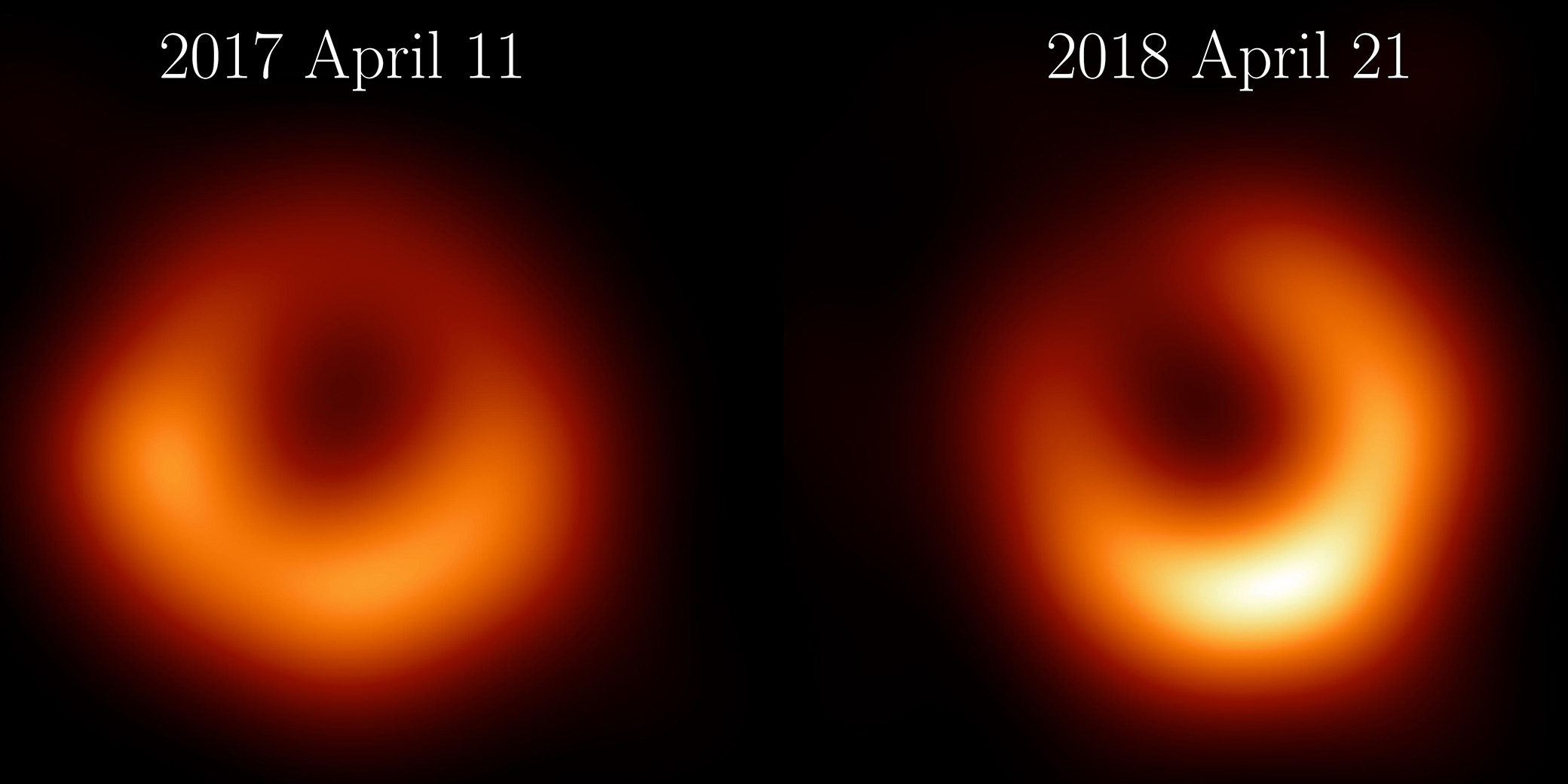}
\includegraphics[width=98mm]{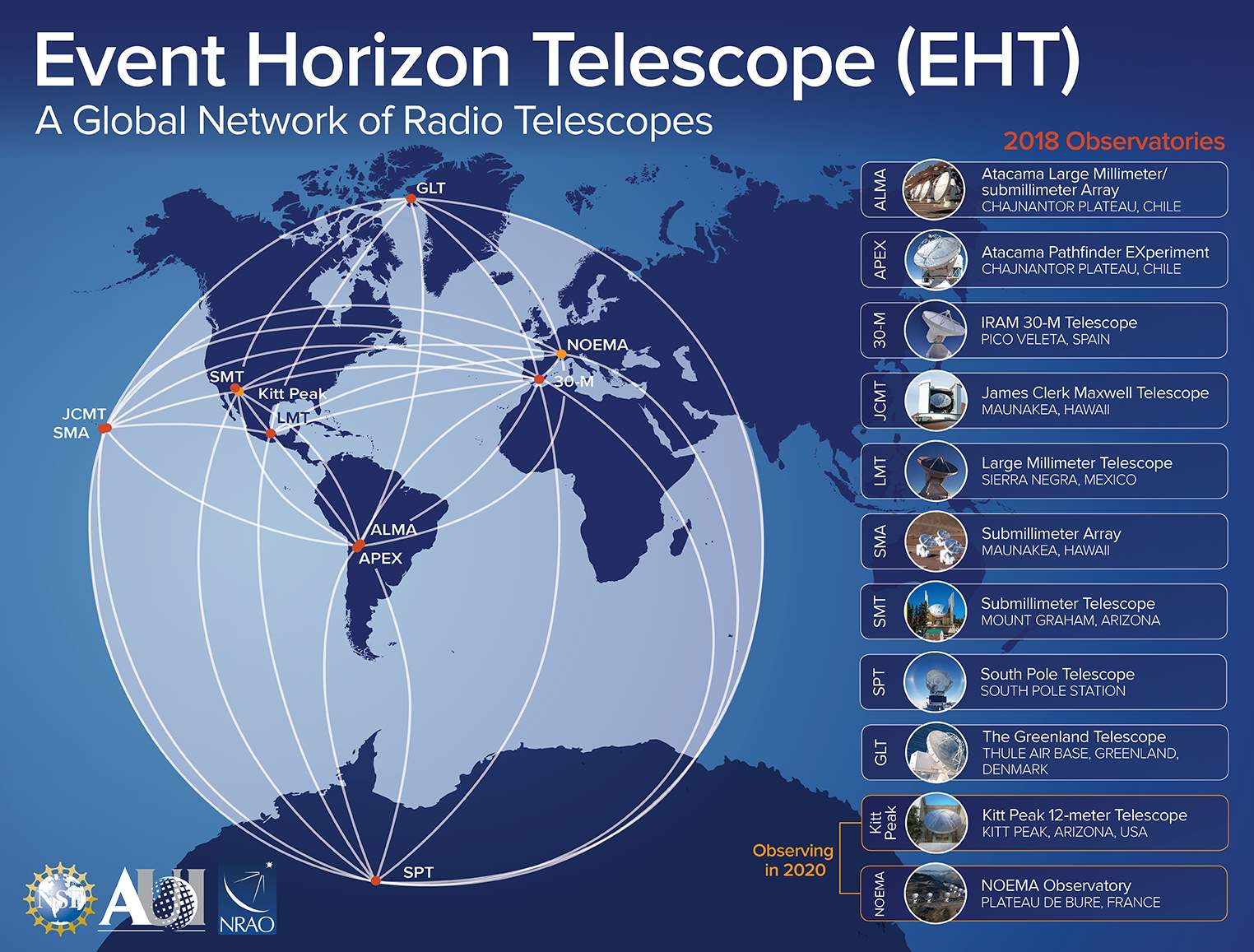}
\caption{(\textbf{Top}) EHT resolution (EHT collaboration~\citep{10.3847/2041-8213/ab0c96}). 
(\textbf{Bottom}) Diagram of the EHT Network used for the observations in 2017. (Credit: Argonne National Laboratory-NRAO/AUI/NSF).}
\label{fig2}
\end{figure}

The EHT collaboration is also developing space-based extensions that would dramatically increase the baseline length and angular resolution~\citep{10.3847/1538-4357/ab86ac, 10.1051/0004-6361/201936622}---see Figure~\ref{fig2}. A~space-based telescope in Earth orbit could deliver baselines exceeding the Earth's diameter, yielding sub-microarcsecond resolution. Such capabilities would allow detailed imaging of jet collimation and acceleration zones, potentially resolving the transition from magnetospheric to hydrodynamic flow regimes~\citep{10.1038/nature07245, 10.3847/1538-4357/aa6193}.

\begin{figure}[H]
\includegraphics[width=99mm]{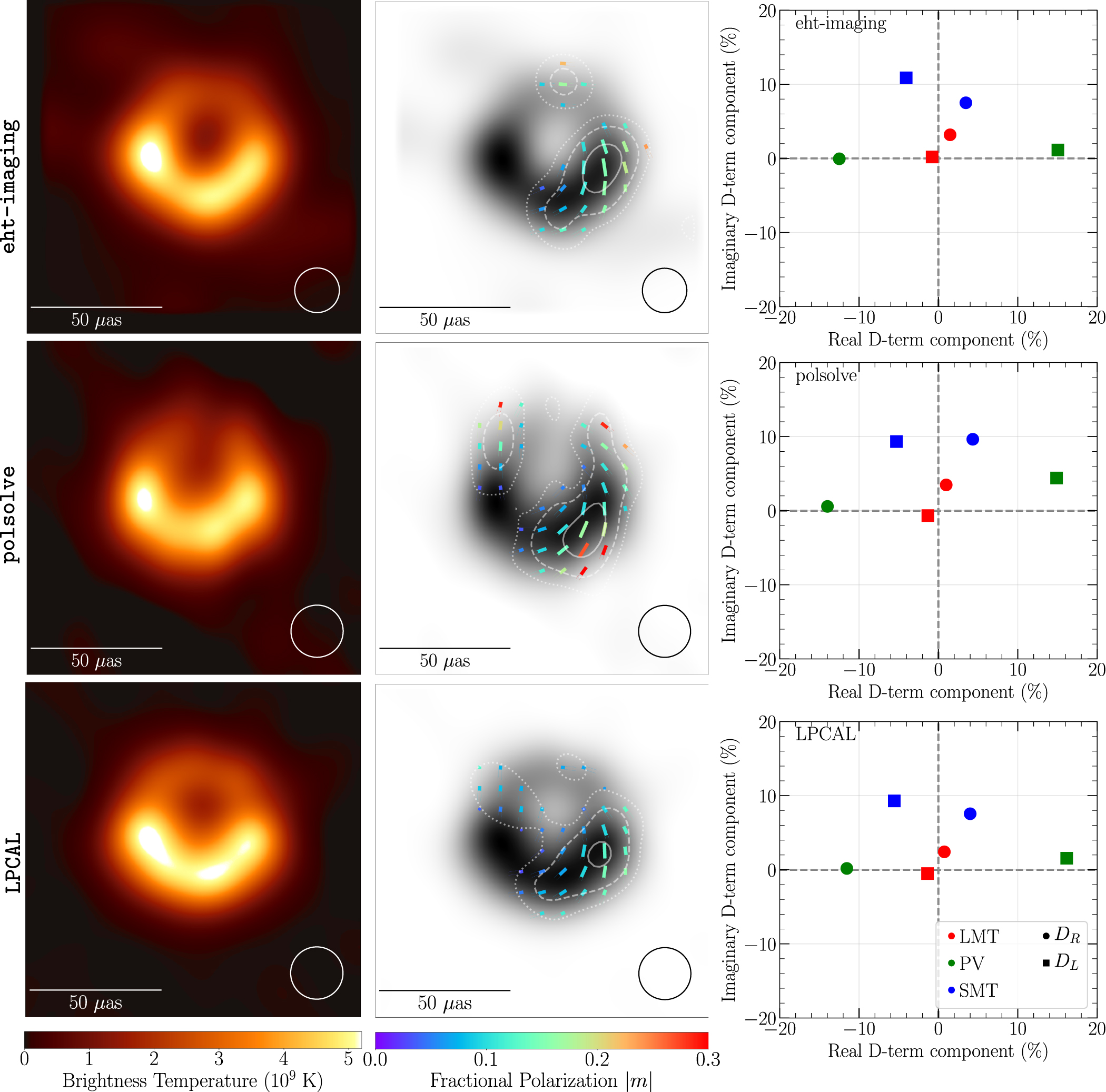}
\vspace{0.32cm}
\caption{Polarimetric imaging of M87* from 2017 April 11 low-band data. (\textbf{Left}) Total intensity images generated by EHT-imaging, polsolve, and~LPCAL. EHT-imaging results are blurred to achieve}
\label{fig3}
\end{figure}

Polarization observations represent another frontier for the EHT~\citep{10.3847/2041-8213/abe71d}. Linear polarization traces the magnetic field structure in the emission region, directly mapping the field geometry responsible for jet launching~\citep{10.1126/science.1094023, 10.1086/510850}---see Figure~\ref{fig3}. Circular polarization can probe the plasma composition and magnetic field strength through Faraday rotation effects~\citep{10.1088/0004-637X/745/2/115, 10.1051/0004-6361/201424358}. The~first polarization observations of M87* have already revealed ordered magnetic field structures consistent with theoretical predictions for magnetically dominated jets~\citep{10.3847/2041-8213/abe71d}.

Multi-frequency polarization observations will facilitate Faraday rotation measure mapping, delimiting constraints on the electron density and magnetic field strength along the line of sight~\citep{10.1088/0004-637X/797/1/66, 10.1142/S0218271817300014}. This technique can distinguish between different models for the plasma composition and magnetic field configuration in the jet launching region. The~combination of total intensity and polarization imaging at multiple frequencies will culminate in a comprehensive view of the physical conditions near the event horizon~\citep{10.1093/mnras/stx587, 10.3847/1538-4357/aab6a8}.

\vspace{-8pt}

\vspace{-6pt}

\vspace{-12pt}
{\captionof*{figure}{a 20 $\upmu$as circular Gaussian resolution, matching polsolve and LPCAL CLEAN images. (\textbf{Middle}) Polarimetric reconstructions following full-array leakage calibration. Total intensity is shown in grayscale background. Polarization ticks represent the Electric Vector Position Angle (EVPA), with~their length proportional to the linear polarization intensity magnitude and color indicating fractional linear polarization. Contours illustrate linear polarized intensity at 20, 10, and~5 $\upmu$Jy $\upmu$as$^{-2}$ (solid, dashed, and~dotted, respectively). Regions with Stokes of peak flux density or less than 20\% of peak polarized flux density were omitted. A~consistent feature across all reconstructions is the predominance of the highest linear polarization fraction and intensity in the southwest part of the ring. (\textbf{Right}) Preliminary D-terms for SMT, PV, and~LMT, derived through the polarimetric leakage calibration method using EHT-imaging, polsolve, and~LPCAL. (Credit: EHT collaboration~\citep{10.3847/2041-8213/abe71d}).}}

\vspace{6pt}
\subsection{The Cherenkov Telescope Array (CTA)}

\begin{figure}[H]
\includegraphics[width=114mm]{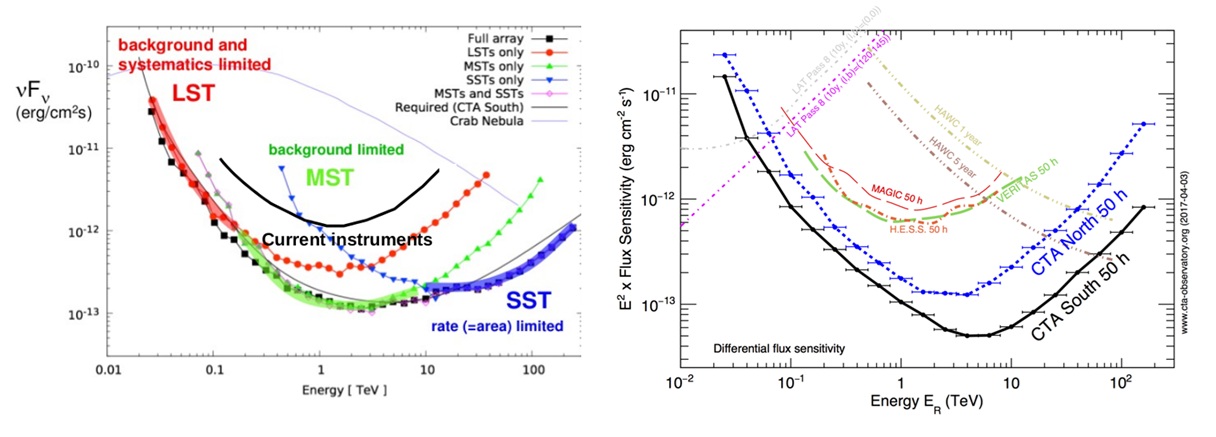}
\includegraphics[width=114mm]{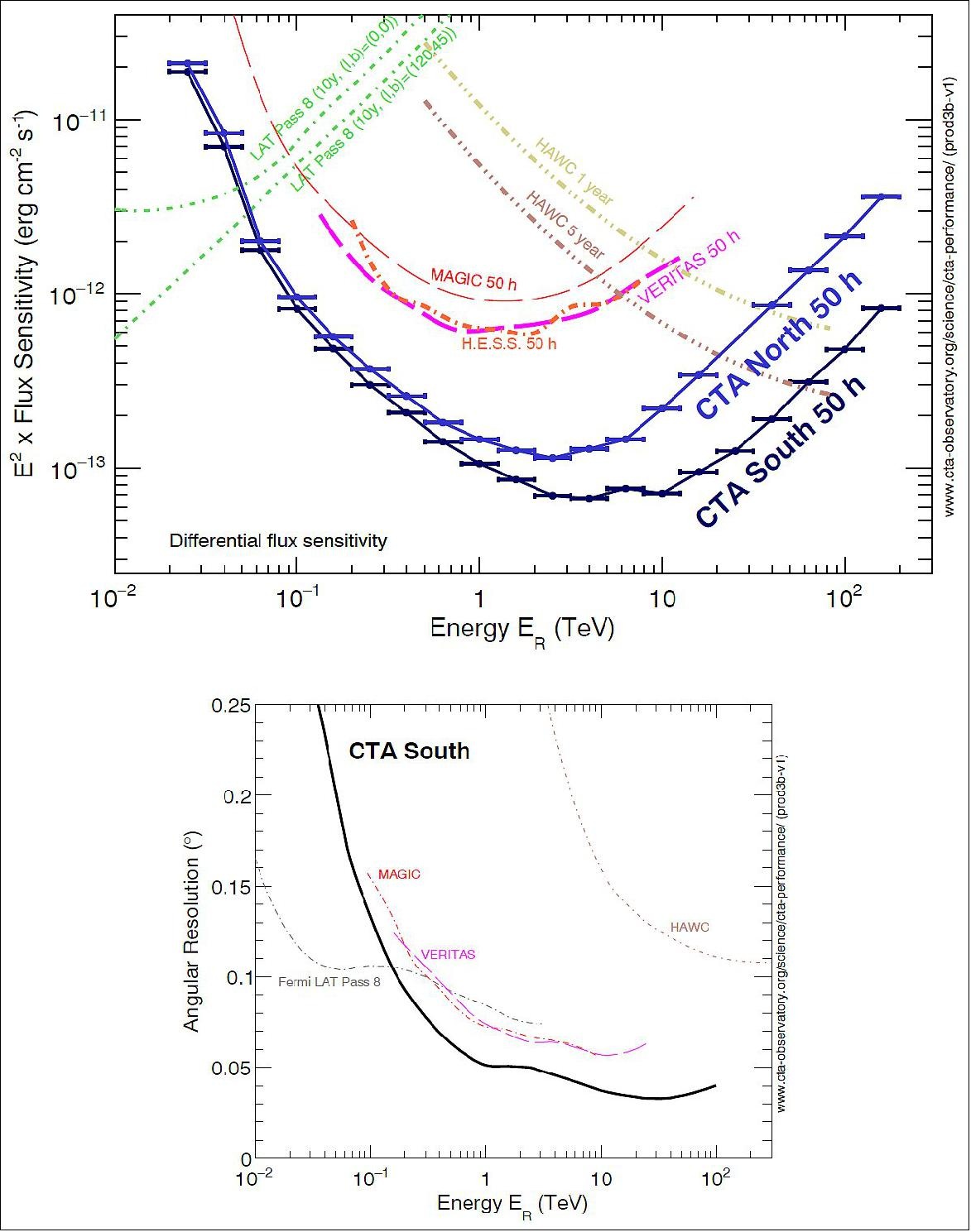}
\caption{\hl{CTA point source} 
 sensitivity comparison. (\textbf{Top}) Overall sensitivity of the full CTA array (black line, filled squares) against individual telescope types: 4 LSTs (red line, filled circles), 25 MSTs (green line, filled triangles), and~75 SSTs (blue line, upside-down triangles). Current instrument sensitivity for the same observation time is also shown (black line), with~expected improvements as analysis algorithms evolve and the final layout is set. (\textbf{Bottom}) Differential flux sensitivity for the CTA southern array (black solid line) and northern array (blue dotted line). For~context, sensitivities of H.E.S.S., VERITAS, MAGIC (for the same observation time), HAWC (one and five-year observations), and~Fermi-LAT (10 years with two diffuse gamma-ray background levels) are included. (Credit: S. Mangano~\citep{10.48550/arXiv.1705.07805} and CTA collaboration-eoPortal~\cite{https://www.eoportal.org/other-space-activities/cta}).}
\label{fig4}
\end{figure}

The CTA will be the most sensitive ground-based gamma-ray observatory to date, covering energies from tens of GeV to hundreds of TeV~\citep{10.1142/10986, 10.1007/s10686-011-9247-0}. Its combination of wide sky coverage, excellent temporal resolution, and~large collecting area makes it ideal for probing the high-energy emission of blazars and radio galaxies. CTA represents a significant advance over current Cherenkov telescopes, with~an order of magnitude improvement in sensitivity and angular resolution~\citep{10.1016/j.astropartphys.2013.01.007, 10.1007/s10686-011-9247-0}---see Figure~\ref{fig4}.

The enhanced capabilities of CTA will empower several key scientific advances in jet physics. Population studies of blazars will distinguish between leptonic and hadronic emission models via the detailed spectral and temporal information for large samples of sources~\citep{10.1088/0004-637X/771/1/54, 10.1093/mnras/sty2636, 10.1088/0004-637X/768/1/54}. The~improved sensitivity will allow detection of fainter sources and extend the redshift range of observable blazars, unveiling insights into the evolution of jet properties with cosmic time~\citep{10.1051/0004-6361:20031615, 10.1088/0004-637X/712/1/238}.

CTA will quantify particle acceleration timescales through high-cadence monitoring of blazar flares~\citep{10.1088/0004-637X/696/2/L150, 10.1086/592348}. The~rapid variability observed in some blazars implies acceleration timescales shorter than the synchrotron cooling time, requiring efficient acceleration mechanisms such as magnetic reconnection or shock acceleration in highly magnetized plasmas~\citep{10.1111/j.1745-3933.2009.00635.x, 10.1088/2041-8205/783/1/L21}. CTA's ability to resolve sub-hour variability will pin down the size and magnetic field strength of acceleration regions~\citep{10.1103/PhysRevD.111.083049, 10.1111/j.1365-2966.2010.18140.x}.

The array's improved angular resolution will enable spatial studies of extended gamma-ray emission from radio galaxy jets~\citep{10.1088/0004-637X/746/2/151, 10.1088/2041-8205/723/2/L207}. Current instruments can barely resolve the jet structure in nearby sources like M87, but~CTA will unveil detailed maps of gamma-ray emission along jets and in radio lobes~\citep{10.1126/science.1134408, 10.1126/science.1175406}. These observations will refine models for particle acceleration and energy dissipation in different regions of the jet, from~the compact core to the extended lobes~\citep{10.1111/j.1365-2966.2006.10525.x, 10.1111/j.1365-2966.2009.14887.x}.

CTA's synergy with neutrino detectors such as IceCube may clarify the contribution of relativistic jets to high-energy cosmic messengers~\citep{10.1126/science.aat2890, 10.1088/1748-0221/12/03/P03012}. Hadronic models for blazar emission predict associated neutrino production through pion decay, but~current limits from IceCube are approaching the predictions of some models~\citep{10.1093/mnras/stac3190, 10.1093/mnras/stv179}. CTA's improved sensitivity to gamma-ray flares will allow for more precise correlation studies with neutrino detections, potentially establishing the first definitive evidence for hadronic acceleration in jets~\citep{10.1093/mnras/sty1852, 10.3847/1538-4357/aaa7ee}.

The wide field of view and rapid slewing capabilities of CTA will ensure follow-up observations of transient events detected by other facilities~\citep{10.1007/s10686-011-9247-0}. Gravitational wave detections of black hole mergers may be associated with jet-powered electromagnetic counterparts, particularly if the merger involves a neutron star~\citep{10.1111/j.1365-2966.2010.16864.x, 10.1093/mnrasl/slx175}. CTA's ability to rapidly respond to alerts and conduct deep observations will be fundamental for detecting such rare events~\citep{10.1088/978-0-7503-1369-8}.

Multi-wavelength coordination with other facilities will maximize the scientific return of CTA observations~\citep{10.1088/0004-637X/768/1/54}. Simultaneous observations with radio telescopes can trace the evolution of synchrotron emission as particles are accelerated to gamma-ray energies~\mbox{\citep{10.1088/2041-8205/710/2/L126, 10.1088/0004-637X/715/1/362}}. X-ray observations furnish complementary information about the electron distribution and magnetic field strength~\citep{10.1086/321394, 10.1051/0004-6361:200810122}. Optical monitoring can characterize external photon fields that contribute to inverse Compton emission~\citep{10.1088/0004-637X/704/1/38, 10.1111/j.1365-2966.2009.15898.x}.

\vspace{-8pt}

\subsection{The Vera C. Rubin Observatory (LSST)}

\begin{figure}[H]
\includegraphics[width=115mm]{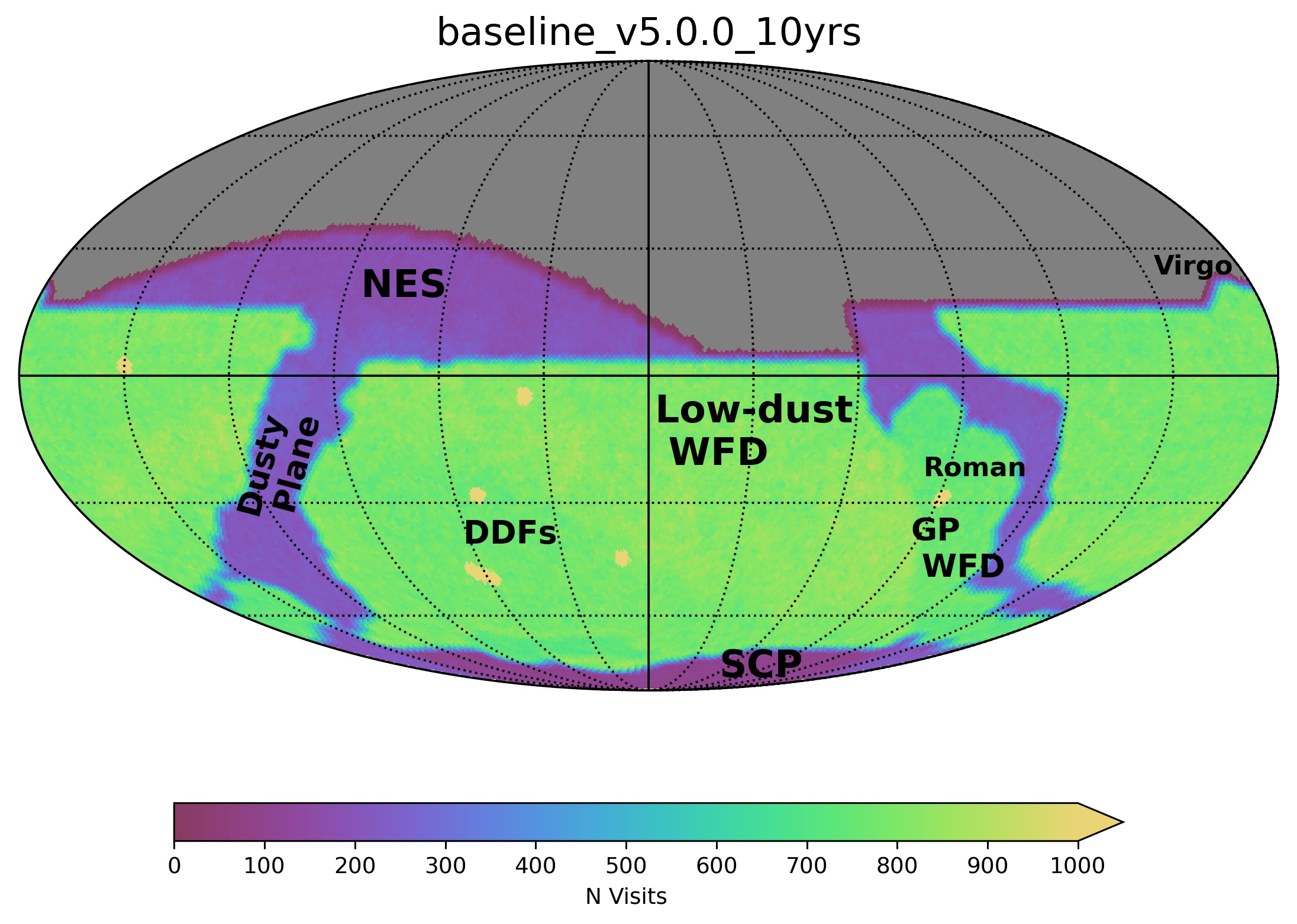}
\caption{\hl{The LSST footprint}
, illustrating the sky observed over the 10-year survey and the distribution of on-sky visits across the major components of the LSST. Where N Visits (number of visits in the LSST footprint) (Credit: Schwamb~et~al. 2023~\citep{10.3847/1538-4365/acc173}).}
\label{fig5}
\end{figure}

The LSST will revolutionize time-domain astrophysics by means of its 10-year survey of the southern sky with unprecedented depth and cadence~\citep{10.48550/arXiv.0912.0201, 10.3847/1538-4357/ab042c}---see Figure~\ref{fig5}. For~AGN and jet studies, LSST will deliver massive statistical samples of optical variability, enabling the study of jet duty cycles, feedback episodes, and~correlations with multi-wavelength activity~\citep{10.1146/annurev-astro-081913-035722, 10.48550/arXiv.1811.06542}. The~survey will observe approximately $2 \times 10^{10}$ galaxies and $1.7 \times 10^{10}$ stars, thereby producing an unprecedented dataset for statistical studies of AGN variability~\citep{10.48550/arXiv.0912.0201}.

\vspace{-8pt}

LSST's observing strategy involves repeated observations of the entire visible southern sky every few nights, with~deeper observations in selected regions~\citep{10.3847/1538-4357/ab042c}. This approach will generate light curves for millions of AGN with typical cadences of 3--4 days and photometric precision of $1\%$ or better for sources brighter than magnitude 24~\citep{10.48550/arXiv.0912.0201}. The~\mbox{10-year} baseline will unlock studies of long-term variability patterns and correlations with host galaxy properties~\citep{10.1088/0004-637X/753/2/106, 10.3847/1538-4357/834/2/111}, and~pave the way for comprehensive studies of AGN duty cycles and triggering mechanisms~\citep{10.1093/mnras/stu327, 10.1088/0004-637X/782/1/9}. 

Current samples of variable AGN are limited by selection effects and small sample sizes, making it difficult to accurately estimate the fraction of galaxies that host active jets and the timescales for jet activity~\citep{10.1111/j.1365-2966.2012.20414.x, 10.1051/0004-6361/201833883}. LSST will furnish unbiased samples of AGN across a wide range of luminosities and redshifts, facilitating statistical studies of jet triggering and its dependence on black hole mass, accretion rate, and~host galaxy properties~\citep{10.1093/mnras/stu2549, 10.1088/0004-637X/701/1/66}.

LSST's ability to capture rare transients, such as tidal disruption events (TDEs) with relativistic jets, will expand our understanding of jet formation beyond the AGN paradigm~\citep{10.1016/j.jheap.2015.04.006, 10.3847/2041-8213/abb6ef}. TDEs represent a unique laboratory for studying jet launching in previously quiescent black holes, potentially revealing the conditions required for jet formation~\citep{10.1093/mnrasl/slw064, 10.3847/1538-4357/ad17c0}. The~survey is expected to discover thousands of TDEs, with~a subset showing evidence for relativistic outflows through radio and X-ray follow-up observations~\citep{10.3847/1538-4357/aa633b, 10.1088/0004-637X/741/2/73}.

The survey will also permit studies of AGN feedback through correlations between jet activity and host galaxy properties~\citep{10.1146/annurev-astro-081811-125521, 10.1038/s41550-018-0403-6}. Optical variability can trace episodes of enhanced accretion and jet activity, which may correlate with star formation rates, galaxy morphology, and~environment~\citep{10.1088/0004-637X/696/1/891, 10.1111/j.1365-2966.2011.19624.x}. Large statistical samples will allow detailed studies of feedback efficiency and its dependence on galaxy mass and \mbox{environment~\citep{10.1111/j.1365-2966.2005.09675.x, 10.1093/mnras/stw2735}}.

LSST's photometric redshift capabilities will make possible studies of AGN evolution across cosmic time~\citep{10.1093/mnras/stz3244, 10.3847/1538-4357/ab042c}. The~survey will detect AGN out to redshifts exceeding 6, shedding new light on insights into the role of jets in early galaxy formation and the growth of the first supermassive black holes~\citep{10.1038/nature25180, 10.1086/344675}. Correlations between jet activity and galaxy assembly can restrict models for AGN feedback in the early universe~\citep{10.1038/nature03335, 10.1086/499298}.

The survey's multi-band photometry will yield crude spectral information for millions of AGN, enabling studies of emission line variability and correlations with continuum changes~\citep{10.1088/0004-637X/753/2/106, 10.3847/1538-4357/834/2/111}. Broad emission lines can respond to continuum variations on timescales of weeks to months, establishing constraints on the size and structure of the broad line region~\citep{10.1086/423269, 10.1088/0004-637X/767/2/149}. Correlations between line and continuum variability can probe the relationship between accretion disk and jet activity~\citep{10.1088/0004-637X/764/1/47, 10.3847/0004-637X/821/1/56}.

\subsection{The Whole Earth Blazar Telescope (WEBT)}

Jorstad~et~al.~\citep{10.1038/s41586-022-05038-9} presents a landmark discovery, rapid ($\sim$13-h) quasi-periodic oscillations simultaneously detected in optical flux, optical polarization, and~$\gamma$-ray flux during a historic outburst of BL Lacertae. The~WEBT’s core achievement here is the collection of an unprecedented 16,497 optical photometric and 1285 polarimetric measurements from 37~telescopes worldwide. This dense, continuous sampling is crucial for identifying the transient, short-period QPOs that would be missed by sparse monitoring.
The strong, zero-lag correlation between optical and $\gamma$-ray fluxes implies a co-spatial origin for the synchrotron (optical) and synchrotron self-Compton ($\gamma$-ray) emission regions. This will directly challenge models where these emissions originate in widely separated zones. By~synergizing WEBT data with monthly 43 GHz VLBA imaging,  the~QPO phenomenon is directly tied to a specific physical structure in the jet. The~oscillations began as a superluminal knot passed through a quasi-stationary recollimation shock, located $\sim$5 pc from the black hole. Also, the~multiwavelength QPOs are modeled as a current-driven kink instability in the jet's helical magnetic field, triggered by the passing shock. This instability creates periodic magnetic reconnection events, accelerating particles and producing the observed correlated flux and polarization oscillations. This model elegantly explains the observed periods and the relationship between flux and polarization~variations.

Long-baseline, high-cadence optical monitoring is essential for triggering and interpreting targeted VLBI and high-energy observations. The WEBT will provide the temporal ``movie'' that gives context to the VLBA ``snapshots'', enabling a direct connection between observed variability and parsec-scale jet~morphology. 

\begin{figure}[H]
\includegraphics[width=115mm]{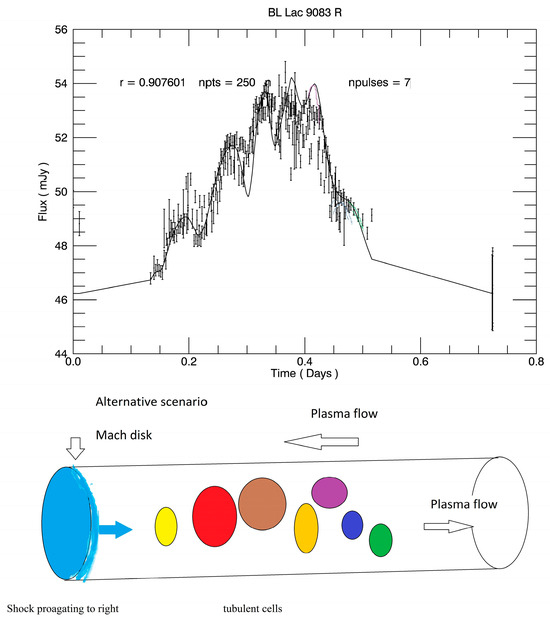}
\caption{\hl{Microvariability curve} 
 of BL Lac (\textbf{top}) alongside an illustrative model (\textbf{bottom}) depicting a shock propagating through a turbulent jet. The~model demonstrates how the shock triggers radiation pulses from colored density enhancements, which subsequently convolve to produce the observed micro-variability curve, i.e., the illustration shows a blue shock moving from left to right along a jet. As it passes each colored region (density enhancement), it causes a burst of radiation. Adding up all these bursts creates the micro-variability plot shown. (Credit: Webb \& Sanz~\citep{10.3390/galaxies11060108}).}
\label{fig6}
\end{figure}

Zooming in from day-long QPOs to minute-hour micro-variability, Webb \& Sanz~\citep{10.3390/galaxies11060108} analyzes the intra-night ``flickering'' in the same 2020 WEBT dataset. The~authors isolate 41 well-sampled nights and fit the micro-variability curves with pulses predicted by the ``turbulent cell'' model.
The authors interpret each micro-variability pulse as synchrotron emission from a single turbulent plasma cell---i.e., a~localized enhancement in density and magnetic field---being energized as it is crossed by a propagating shock. By~deconvolving the observed light curves, they infer the characteristic sizes of these turbulent cells.
They probe the micro-physics of the jet plasma---the very medium in which the large-scale kink instability develops---see Figure~\ref{fig6}.
The inferred cell sizes (order of AU to tens of AU) inform the inner boundary conditions for simulations of jet turbulence and instability growth. The~excellent fits of the turbulent cell model (average correlation coefficient of 0.948) to numerous independent nights strongly support the picture that the blazar jet flow is fundamentally turbulent. This turbulence is likely the source of the ``random'' magnetic field component that dilutes polarization.
The WEBT's unique capability---coordinated, high-density, long-duration, multi-observer monitoring---has proven indispensable. It will bridge the gap between sporadic space-telescope observations and infrequent VLBI imaging, allowing researchers to connect the dots across time, frequency, and~physical scale. The 2020 campaign on BL Lacertae stands as a paradigm for how such collaborations can transform blazar variability from a chaotic curiosity into a powerful diagnostic of relativistic jet physics, from~global MHD instabilities down to micro-parsec turbulent plasma structures. the WEBT observations are the ultimate testing ground for the turbulent cell/kink instability~model.

\vspace{-8pt}
\unskip

\subsection{Synergies and Multi-Messenger~Astronomy}

Before delving into the synergies and multi-messenger astronomy, it is useful to consult Table~\ref{tab:observatories}, which presents a comparative overview of the Next-Generation Observatories for Black Hole Jet~Studies.

The true power of these facilities lies in their complementarity and the synergies that emerge from coordinated observations. Horizon-scale imaging by the EHT, high-energy light curves from CTA, and~long-term monitoring by LSST together provide a multi-scale view of jet physics, from~microphysical processes near the event horizon to kiloparsec-scale propagation~\citep{10.1146/annurev-astro-081817-051948}. These observational advances, combined with X-ray and neutrino facilities, herald a new era of multi-messenger jet astrophysics~\citep{10.1038/s42254-019-0101-z}.

\begin{table}[H]
\caption{Comparative Table of Next-Generation Observatories for Black Hole Jet~Studies.}
\label{tab:observatories}
\footnotesize
\begin{adjustwidth}{-\extralength}{0cm}
		\begin{tabularx}{\fulllength}{LLLLLLLL}
			\toprule

%
    \textbf{Observatory} & \textbf{Wavelength/} \textbf{Energy Range} & \textbf{Angular\linebreak Resolution (or Scale)} & \textbf{Sensitivity} \textbf{(Flux Limit)} & \textbf{Sky Coverage/\linebreak Field of View} & \textbf{Temporal\linebreak Cadence} & \textbf{Key\linebreak Contributions to Jet Physics} \\
    \hline
    \textbf{\hl{EHT}} 
 (Event Horizon Telescope) & Millimeter waves (e.g., 1.3 mm, 0.87 mm) & Horizon scales ($\upmu$as) & High (for bright, nearby objects) & Pointed targets (Sgr A*, M87*) & Rapid variability (minutes\ seconds) & Direct imaging of jet launching regions, magnetic field structures, GR~tests \\
    \hline
    \textbf{\hl{CTA}} 
 (Cherenkov Telescope Array) & Gamma-rays (GeV-TeV) & Significant angular improvement (arcmin) & $\sim$1 order of magnitude better than current & Wide sky coverage, rapid slewing & Blazar variability (sub-hour) & Particle acceleration, emission models, transient event detection, neutrino~correlation \\
    \hline
    \textbf{\hl{LSST}} 
 (Vera C. Rubin Observatory) & Optical (u, g, r, i, z, y bands) & Wide, for~galactic surveys & Down to 27.5~mag & Half sky (Southern Hemisphere) & High cadence (few nights) & AGN variability, jet duty cycles, feedback, TDEs with jets, statistical~studies \\
    \hline
    \textbf{\hl{WEBT}} 
 (Whole Earth Blazar Telescope) & Multi-wavelength Optical & No intrinsic spatial resolution (point-like) & High (for blazar monitoring) & Selected point targets & Dense, continuous monitoring (h/days) & Rapid variability (QPOs, micro-variability), turbulent plasma physics, kink~instability \\

	\bottomrule
		\end{tabularx}
	\end{adjustwidth}
\end{table}

Coordinated observations between EHT and CTA can directly connect horizon-scale jet launching with high-energy emission~\citep{10.3847/1538-4357/abac0d}. Simultaneous observations during gamma-ray flares can test whether variability originates in the jet launching region or further downstream~\citep{10.1088/2041-8205/710/2/L126}. The~time delays between millimeter and gamma-ray variations can trace the location of gamma-ray emission and the propagation speed of disturbances along the jet~\citep{10.1093/mnras/stu1749, 10.1093/mnras/stu540}.

LSST's wide-area surveys will identify targets for detailed follow-up with EHT and CTA~\citep{10.48550/arXiv.1811.06542}. The~survey will discover new classes of variable AGN and transients that may be associated with jet activity~\citep{10.1093/mnras/stz3244}. Rapid alerts from LSST can trigger target-of-opportunity observations with other facilities, thereby allowing multi-wavelength studies of jet flares and outbursts~\citep{10.1088/1538-3873/aaecbe}.

The WEBT collaboration has provided the most comprehensive, high-cadence, multi-wavelength datasets on blazar variability. When analyzed through the lens of modern theoretical frameworks, these observations do more than just document phenomena—they allow us to deconstruct the physics of relativistic jets across all scales~\citep{10.1038/s41586-022-05038-9, 10.3390/galaxies11060108}.

The combination of electromagnetic and gravitational wave observations opens new possibilities for studying jet physics~\citep{10.3847/2041-8205/821/1/L18}. Black hole mergers detected by LIGO/Virgo may be associated with electromagnetic counterparts powered by jets, particularly in asymmetric mergers or mergers involving neutron stars~\citep{10.1111/j.1365-2966.2010.16864.x, 10.1093/mnrasl/slx175}. Multi-messenger observations can determine the efficiency of jet launching in merger events and the role of magnetic fields in the merger process~\citep{10.3847/2041-8205/824/1/L6, 10.1103/PhysRevD.95.063016}.

Neutrino astronomy represents another complementary probe of jet physics~\citep{10.1126/science.aat2890}. High-energy neutrinos can only be produced through hadronic processes, constituting a direct test of hadronic acceleration models in jets~\citep{10.1093/mnras/stac3190}. The~correlation of neutrino detections with gamma-ray flares observed by CTA can distinguish between leptonic and hadronic emission mechanisms and quantify the cosmic ray acceleration efficiency of jets~\citep{10.1093/mnras/sty1852, 10.3847/1538-4357/aaa7ee}.

Future space-based gravitational wave detectors such as LISA (Laser Interferometer Space Antenna) will extend the sensitivity to lower frequencies, potentially detecting the inspiral phase of massive black hole mergers~\citep{10.48550/arXiv.1702.00786}. These events may be associated with electromagnetic precursors as the black holes approach merger, offering advance warning for electromagnetic follow-up observations~\citep{10.1103/PhysRevD.96.023004}. The~combination of gravitational wave and electromagnetic observations can uncover the role of magnetic fields and jets in the merger process~\citep{10.1103/PhysRevLett.111.061105}.

The development of real-time data processing and alert systems will be decisive for maximizing the scientific return of multi-messenger observations~\citep{10.1088/1538-3873/aaecbe}. Rapid identification and characterization of transient events will ensure coordinated follow-up observations across multiple wavelengths and messengers. Machine learning techniques will play an increasingly important role in identifying interesting events in the massive data streams from these facilities~\citep{10.48550/arXiv.1904.07248, 10.1086/668468}.

\section{Jet Diversity Across Astrophysical~Systems}\label{sec3}

\begin{figure}[H]
\includegraphics[width=115mm]{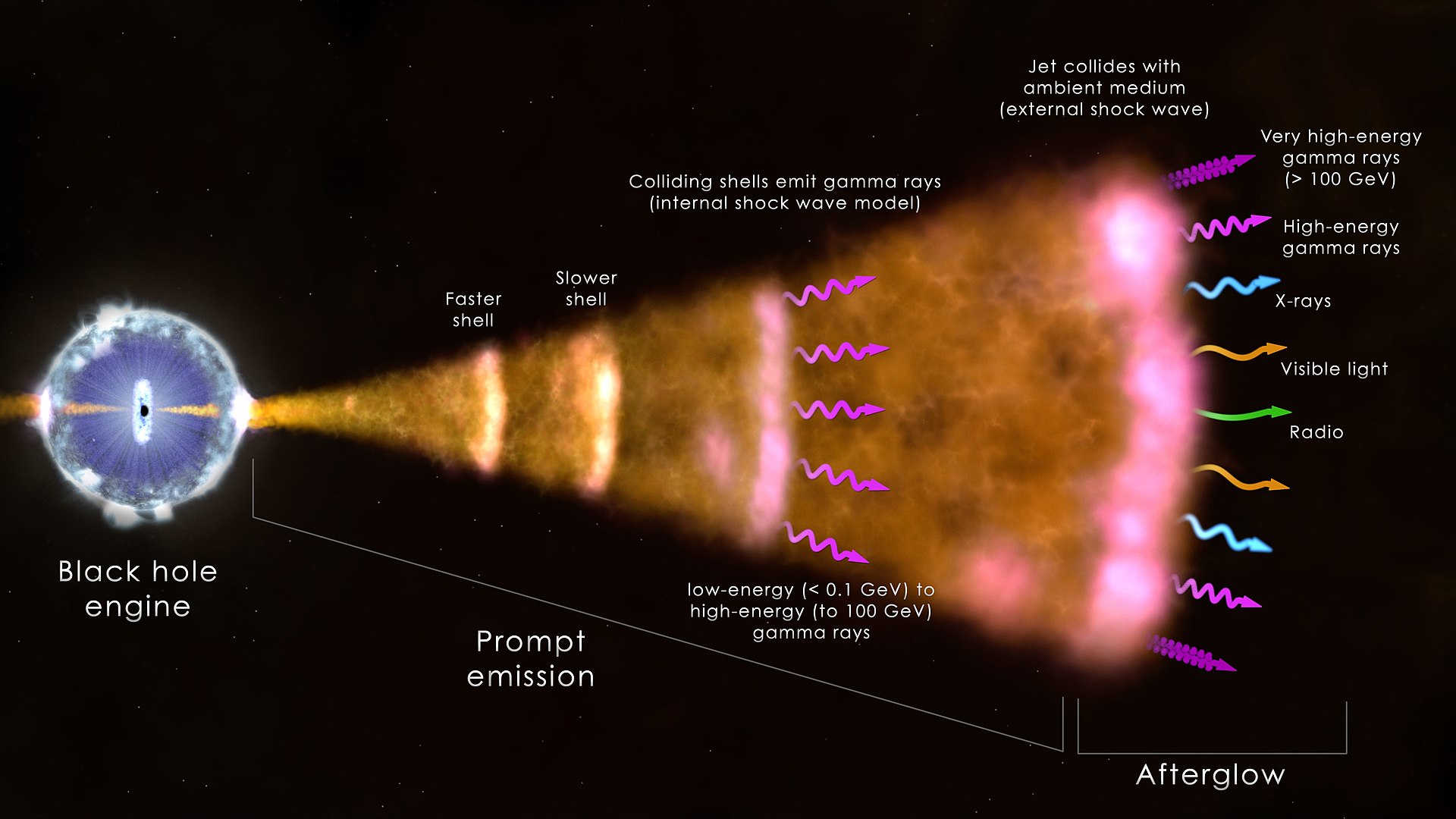}
\caption{Illustration that shows how the jets of a black hole eject enery ranging from gamma rays to
radio frequencies (Credit: NASA/Goddard Research Center/ICRAR).}
\label{fig7}
\end{figure}

The observational capabilities described in the previous section will pave the way for studies of relativistic jets across a wide range of astrophysical systems. While this review focuses primarily on jets from supermassive black holes, the~universality of jet physics can be tested by comparing different mass scales and environments. This diversity of jet phenomena demands important tests of theoretical models and motivates the development of unified frameworks that can explain jet physics across all mass scales and environments---see Figure~\ref{fig7}.

\unskip

\subsection{Stellar-Mass Black Holes and~Microquasars}

X-ray binaries containing stellar-mass black holes serve as laboratories for studying jet physics on much shorter timescales than their supermassive counterparts~\citep{10.1111/j.1365-2966.2009.14841.x, 10.1051/0004-6361:20042457}. The~characteristic variability timescales scale with black hole mass as $t \propto M_{\textrm{BH}}$, allowing observations of complete outburst cycles and state transitions on human \mbox{timescales~\citep{10.1146/annurev.astro.44.051905.092532, 10.1126/science.1221790}.}

Microquasars such as GRS 1915+105 and Cygnus X-1 exhibit jet launching and quenching on timescales of minutes to hours, unveiling real-time observations of the jet formation process~\citep{10.1038/371046a0, 10.1038/nature03879}. These systems have revealed the existence of discrete jet ejection events and the correlation between accretion state and jet properties~\citep{10.1111/j.1365-2966.2004.08384.x, 10.1093/mnras/staa476}.

The scaling relationships between stellar-mass and supermassive black holes constitute important tests of jet physics~\citep{10.1046/j.1365-2966.2003.07017.x, 2007ASPC..373.....H}. The~fundamental plane of black hole activity, relating radio luminosity $L_R$, X-ray luminosity $L_X$, and~black hole mass $M_{\textrm{BH}}$, \hl{follows the~relation:} 
\begin{equation}
\log L_R = \xi_R \log L_X + \xi_M \log M_{\textrm{BH}} + \beta
\end{equation}
where $\xi_R \approx 0.6$, $\xi_M \approx 0.8$, and~$\beta$ is a normalization constant~\citep{10.1088/0004-637X/773/1/59, 10.3847/1538-4357/aaf6b9}. This suggests universal jet launching mechanisms across the mass~scale.

\subsection{Gamma-Ray Bursts as Extreme~Jets}

Gamma-ray bursts (GRBs) represent the most extreme manifestation of relativistic jets, with~Lorentz factors potentially exceeding $\Gamma > 10^3$~\citep{10.1103/RevModPhys.76.1143, 10.1016/j.physrep.2014.09.008}. Long GRBs are associated with the collapse of massive stars, while short GRBs likely originate from neutron star mergers~\citep{10.1146/annurev.astro.43.072103.150558, 10.1146/annurev-astro-081913-035926}.

The detection of GW170817 and its associated short GRB represented the first direct confirmation of jet formation in neutron star mergers~\citep{10.1103/PhysRevLett.119.161101, 10.3847/2041-8213/aa8f41}. The~structured jet model, with~a narrow ultra-relativistic core surrounded by a wider, slower component, has emerged as a leading explanation for the observed properties~\citep{10.1103/PhysRevLett.120.241103, 10.1038/s41586-018-0486-3}. The~jet structure can be parameterized as:
\begin{equation}
\Gamma(\theta) = \Gamma_c + (\Gamma_s - \Gamma_c) \left(\frac{\theta}{\theta_c}\right)^{-k}
\end{equation}
where $\Gamma_c$ is the core Lorentz factor, $\Gamma_s$ is the sheath Lorentz factor, $\theta_c$ is the core half-opening angle, and~$k$ is the power-law~index.

GRB jets open a window onto the early universe, as~they can be detected out to redshifts exceeding $z > 8$~\citep{10.1038/nature08459, 10.1088/0004-637X/736/1/7}. The~use of GRBs as cosmological probes and tracers of star formation in the early universe represents an important application of jet physics~\mbox{\citep{10.1088/0004-637X/749/1/68, 10.3847/0004-637X/817/1/7}.}

\subsection{Tidal Disruption Events with~Jets}

Tidal disruption events (TDEs) occur when a star passes within the tidal radius of a supermassive black hole:
\begin{equation}
r_t = R_* \left(\frac{M_{\textrm{BH}}}{M_*}\right)^{1/3}
\end{equation}
where $R_*$ and $M_*$ are the stellar radius and mass, respectively~\citep{10.1038/333523a0, 10.1017/S0074180900187054}. A~small fraction of TDEs ($\sim$1\%) launch relativistic jets, providing insights into jet formation in previously quiescent black holes~\citep{10.1126/science.1207150, 10.1038/nature10374}.

The jet launching efficiency in TDEs appears to depend on the black hole spin and the magnetic field threading the disrupted stellar material~\citep{10.1093/mnrasl/slw064, 10.3847/1538-4357/ad17c0}. The~magnetic flux required for jet formation can be estimated as:
\begin{equation}
\Phi \gtrsim \sqrt{\dot{M} r_g\;c} \sqrt{4\pi}
\end{equation}
where $\dot{M}$ is the accretion rate and $r_g = GM_{\textrm{BH}}/c^2$ is the gravitational~radius.

\subsection{Scaling Relations and Universal~Properties}

The diversity of jet-producing systems reveals several universal scaling relations that restrict theoretical models. The~jet power scales with accretion rate and black hole mass as:
\begin{equation}
P_{\textrm{jet}} \propto \dot{M}^{\alpha} M_{\textrm{BH}}^{\beta}
\end{equation}
where observations suggest $\alpha \approx$  1.2--1.4 and $\beta \approx $ 0.8--1.2~\citep{10.1126/science.1227416}.

The jet opening angle appears to be inversely correlated with jet power:
\begin{equation}
\theta_j \propto P_{\textrm{jet}}^{-\gamma}
\end{equation}
with $\gamma \approx $ 0.1--0.3~\citep{10.1051/0004-6361/200913422, 10.3847/1538-4357/aa8407}. This relation may reflect the role of magnetic collimation in powerful~jets.

The maximum jet Lorentz factor scales with the dimensionless black hole spin parameter $a_*$ as:
\begin{equation}
\Gamma_{\max} \propto \left(\frac{a_*}{1 + \sqrt{1-a_*^2}}\right)^{1/2}
\end{equation}
consistent with the Blandford-Znajek mechanism~\citep{10.1111/j.1745-3933.2011.01147.x, 10.1111/j.1365-2966.2012.21074.x}.

This diversity of jet phenomena across different astrophysical systems compels fundamental tests of theoretical models and motivates the development of unified frameworks that can explain jet physics across all mass scales and environments. The~scaling relations observed across different systems suggest that the fundamental physics of jet launching and propagation is universal, despite the vastly different environments in which jets~operate.

\section{Theoretical Advances and~Challenges}\label{sec4}

The theoretical understanding of black hole jets has advanced significantly over the past decades, driven by improvements in computational capabilities and the development of sophisticated numerical models. However, the~multi-scale and multi-physics nature of jet phenomena presents ongoing challenges that require innovative approaches combining fluid dynamics, plasma physics, and~general relativity. This section discusses the current state of theoretical models and the key challenges that must be addressed to fully understand jet~physics.

\subsection{GRMHD Simulations and Jet~Launching}

General relativistic magnetohydrodynamic (GRMHD) simulations have established the Blandford-Znajek process as a leading mechanism for jet production~\citep{10.1093/mnras/179.3.433, 10.1088/0004-637X/711/1/50, 10.1111/j.1365-2966.2012.21074.x}. These simulations solve the coupled Einstein-Maxwell-MHD equations in the strong gravitational field near black holes, creating self-consistent models for the extraction of rotational energy from spinning black holes through magnetic field lines threading the ergosphere~\citep{10.1111/j.1365-2966.2004.07738.x, 10.1086/374594}.

Recent works demonstrate that MAD configurations can tap black hole spin energy with high efficiency, producing jets that rival accretion luminosity~\citep{10.1093/mnras/stac285, 10.1016/j.newast.2010.03.001}. In~the MAD state, the~magnetic flux threading the black hole reaches saturation levels, leading to efficient energy extraction through the Blandford-Znajek mechanism~\citep{10.1086/375769, 10.1111/j.1365-2966.2006.10256.x}. The~jet power in this regime scales approximately as:
\begin{equation}
P_{\textrm{jet}} \propto a_*^2 \Phi_{\textrm{BH}}^2 c
\end{equation}
where $a_*$ is the dimensionless black hole spin, $\Phi_{\textrm{BH}}$ is the magnetic flux threading the black hole, and~$c$ is the speed of light~\citep{10.1111/j.1745-3933.2011.01147.x}.

The magnetic flux saturation condition can be expressed as:
\begin{equation}
\Phi_{\textrm{BH}} \lesssim \Phi_{\textrm{MAD}} \equiv \sqrt{\dot{M} r_g\;c} \sqrt{4\pi}
\end{equation}
where $\dot{M}$ is the mass accretion rate and $r_g = GM_{\textrm{BH}}/c^2$ is the gravitational~radius.

\begin{figure}[H]
\includegraphics[width=115mm]{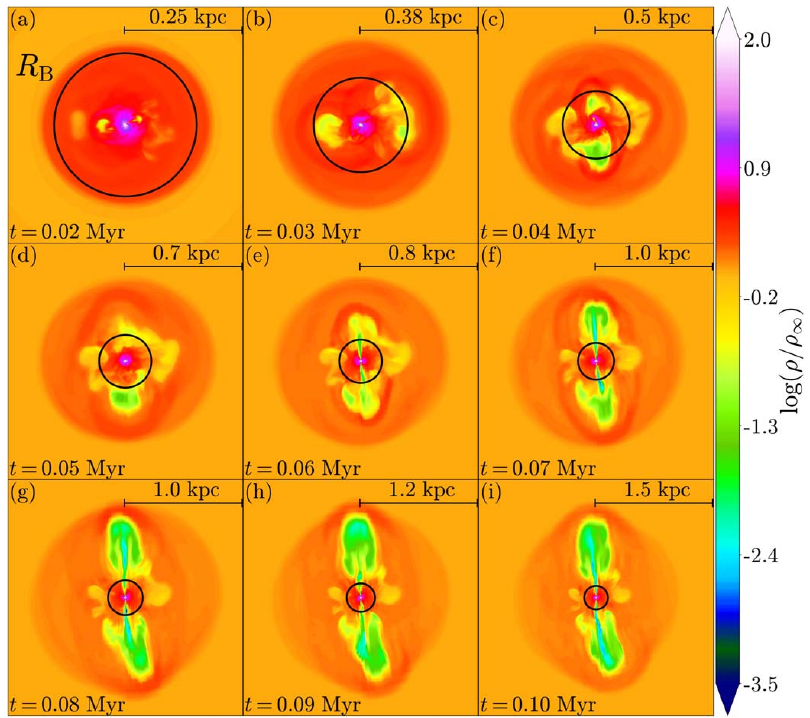}
\caption{\hl{First 3D GRMHD simulation} 
 demonstrating the natural emergence of X-shaped radio galaxy morphology from initially axisymmetric conditions. Panels (\textbf{a}--\textbf{i}) display a time sequence of vertical density slices, scaled to M87*, where the color bar depicts the jet density variation. Initially intermittent, low-density jets (t $\lesssim$ 0.05 Myr) are frequently disrupted but extend beydond the Bondi radius. As~the black hole saturates with magnetic flux and enters a MAD state (t $\sim$ 0.05 Myr), jets stabilize, propagate along the z-axis, explaining the rarity of XRGs. This simulation achieves the largest scale separation ($R_B/R_g = 10^3$) to date (Lalakos~et~al. (2022)~\citep{10.3847/2041-8213/ac7bed}).}
\label{fig8}
\end{figure}

\begin{figure}[H]
\includegraphics[width=115mm]{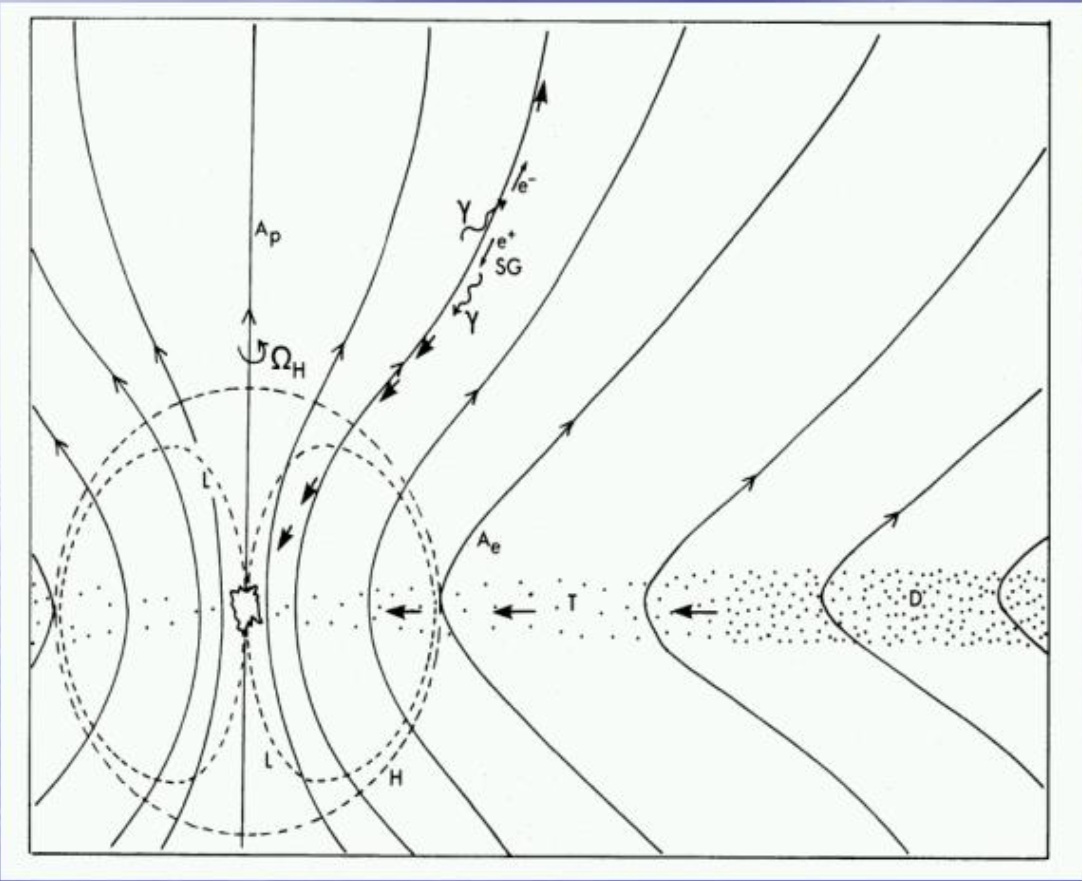}
\caption{\hl{Blandford-Znajek} 
 mechanism. (Credit: Blandford-Znajek (1977)~\citep{10.1093/mnras/179.3.433}).}
\label{fig9}
\end{figure}

GRMHD simulations have revealed several key insights into jet launching\linebreak mechanisms---see Figure~\ref{fig8}. The~magnetic field configuration plays a fundamental role in determining jet properties, with~large-scale poloidal fields being most efficient for energy extraction~\citep{10.1086/533492, 10.1093/mnras/stv014}. The~simulations show that jets naturally develop a spine-sheath structure, with~a highly relativistic spine surrounded by a slower, more massive sheath~\citep{10.1111/j.1365-2966.2006.10256.x, 10.1111/j.1365-2966.2012.22002.x}. This structure arises from the radial dependence of the magnetic field strength and the varying efficiency of the Blandford-Znajek mechanism across the jet cross-section---see Figure~\ref{fig9}.

The simulations also demonstrate that jet power and variability are intimately connected to the accretion flow properties~\citep{10.1093/mnrasl/slx174, 10.3847/2041-8213/ac46a1}. In~the MAD state, periodic magnetic flux eruptions occur when the accumulated magnetic flux exceeds the maximum sustainable level~\citep{10.1086/529025, 10.1016/j.newast.2010.03.001}. These eruptions temporarily reduce the jet power and may explain the observed variability in AGN and X-ray binaries~\citep{10.1093/mnras/stae860, 10.1111/j.1365-2966.2012.20409.x}. The~timescale for these eruptions is typically several times the orbital period at the innermost stable circular orbit:
\begin{equation}
t_{\text{eruption}} \sim (3-10) \times \frac{2\pi r_{\text{ISCO}}}{v_{\text{ISCO}}}
\end{equation}
where $r_{\text{ISCO}}$ and $v_{\text{ISCO}}$ are the radius and velocity at the innermost stable circular~orbit.

However, such simulations often assume idealized plasma conditions, leaving open questions about microphysics and radiative processes~\citep{10.1093/mnras/stv2084, 10.3847/0004-637X/829/1/11}. The~MHD approximation assumes that the plasma can be treated as a conducting fluid, neglecting kinetic effects that become important on small scales~\citep{10.1103/PhysRevLett.106.195003, 10.1093/mnras/stx2530}. The~simulations typically use simple prescriptions for the equation of state and neglect radiative cooling, which can significantly affect the dynamics in luminous sources~\citep{10.1093/mnras/stt1881, 10.3847/1538-4357/ab29ff}.

Current GRMHD codes face several technical challenges that limit their applicability to realistic astrophysical systems~\citep{10.3847/0067-0049/225/2/22, 10.3847/1538-4365/ac9966}. The~numerical resolution required to capture both the horizon scale physics and the large-scale jet propagation exceeds current computational capabilities. Most simulations are limited to relatively small computational domains, typically extending only a few hundred gravitational radii from the black hole~\citep{10.1186/s40668-017-0020-2, 10.1051/0004-6361/201935559}. This limitation prevents the study of jet collimation and interaction with the external~medium.

The treatment of magnetic field evolution in GRMHD simulations remains challenging due to the need to maintain the divergence-free condition $\nabla \cdot \mathbf{B} = 0$~\citep{10.1006/jcph.2000.6519, 10.1006/jcph.2001.6961}. Various numerical schemes have been developed to address this issue, including constrained transport methods and divergence cleaning techniques~\citep{10.1086/308344, 10.1088/0067-0049/198/1/7}. However, the~accumulation of numerical errors over long simulation times can still lead to unphysical magnetic field~configurations.

Recent advances in GRMHD simulations include the development of adaptive mesh refinement techniques that allow higher resolution in regions of interest while maintaining computational efficiency~\citep{10.3847/0067-0049/225/2/22, 10.3847/1538-4365/ab3922}. These methods bridge the study of multi-scale phenomena, from~the event horizon to the jet propagation region. The~implementation of more sophisticated microphysics, including electron heating and cooling processes, is also an active area of development~\citep{10.3847/2041-8213/ab9532, 10.3847/0004-637X/829/1/11}.

It is also worth mentioning the work of Brian Punsly~\citep{10.1086/178134, 10.1086/178135, 10.1086/303471} as a refinement of the Blandford-Znajek mechanism, in~which he embedded it within a more complete and physically realistic system.
Punsly (i) corrected a key dynamical oversight: Magnetic flux dynamics force the strongest field into the ergosphere, not onto the horizon; (ii) identified a more powerful engine: For the fastest-spinning holes, the ergospheric disk wind is the primary power source, not the horizon wind; (iii) proposed a dual-jet structure: This naturally creates a stable, powerful sheath (from ergosphere) plus a variable, energetic core (from horizon); (iv) created a unified diagnostic model: The relative power of these two jets, dictated by black hole spin and accretion history, determines whether we see an OVV quasar, a~BL Lac, or~a hybrid source.
In essence, Punsly's work argued that the Blandford-Znajek jet is the ``blazar within the quasar''. It is the inner, violent heart of the engine, whose relative visibility is controlled by the strength of the more powerful but steadier ergospheric sheath jet that surrounds it. This framework remains a profound and influential way to understand the diversity of relativistic~jets.

\subsection{Plasma Microphysics and Particle~Acceleration}

Jets are not simply fluid structures but complex plasma environments where shocks, turbulence, and~magnetic reconnection can accelerate particles to ultra-relativistic energies~\citep{10.1088/2041-8205/783/1/L21, 10.1088/0004-637X/809/1/38}. Understanding these microphysical processes is necessary for connecting GRMHD simulations with observational signatures, as~the emission properties depend critically on the particle distribution function and magnetic field structure on small scales~\citep{10.1093/mnras/stu2364, 10.1063/1.3703318}.

\begin{figure}[H]
\includegraphics[width=115mm]{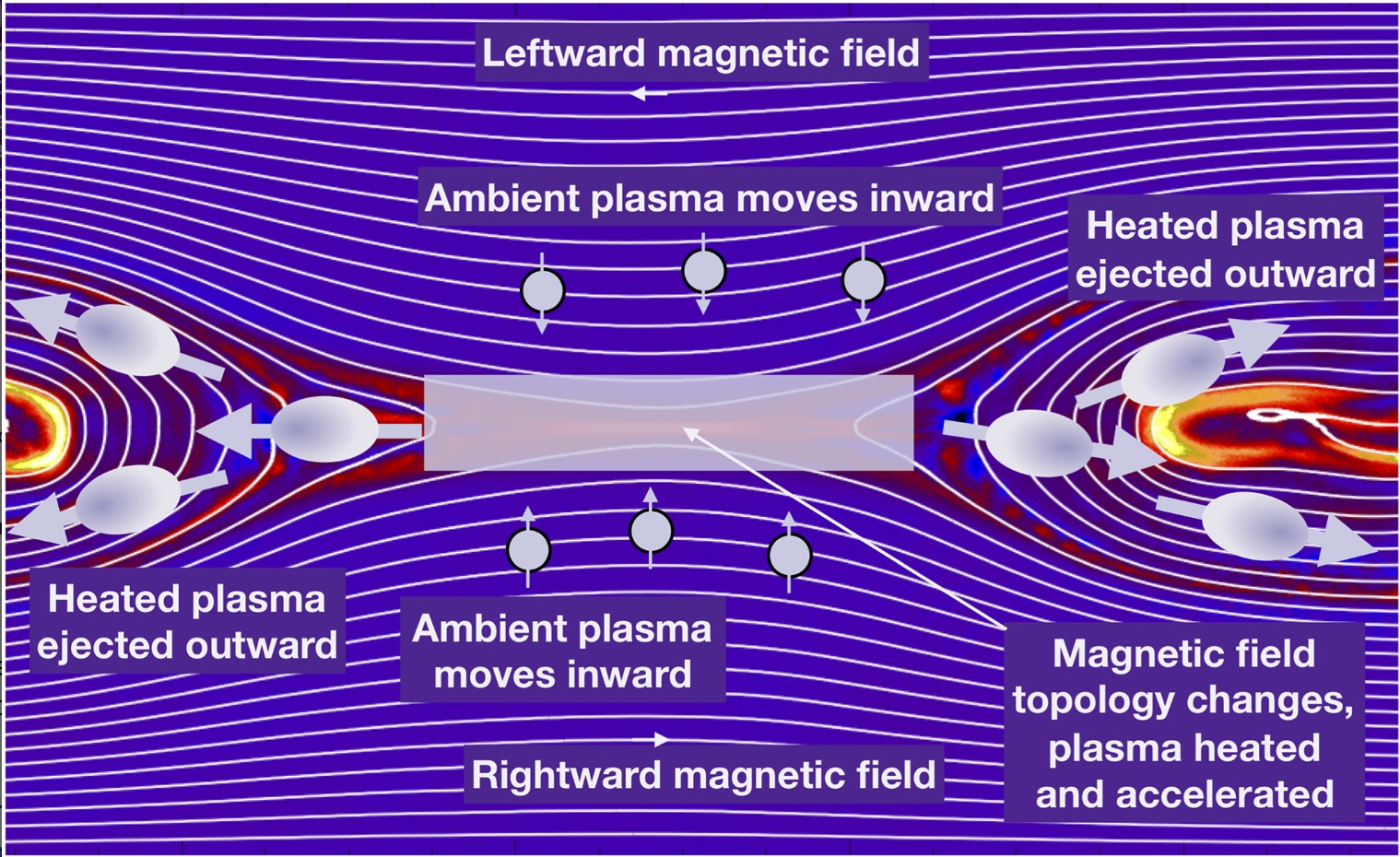}
\caption{Simplified two-dimensional schematic of magnetic reconnection. Opposing magnetic fields (light blue lines) and ambient plasma (light blue circles) converge into a central diffusion region (shaded box), where reconnection occurs. This process heats and accelerates the plasma, ejecting it into jets (shaded blue ovals) to the left and right. (Credit: Hesse \& Cassak (2019)~\citep{10.1029/2018JA025935}).}
\label{fig10}
\end{figure}

Particle-in-cell (PIC) simulations have revealed that reconnection can efficiently produce nonthermal power-law particle distributions consistent with observed blazar spectra~\citep{10.1088/0004-637X/771/1/54, 10.1103/PhysRevLett.113.155005}. Magnetic reconnection occurs when oppositely directed magnetic field lines come into contact and reconnect, releasing magnetic energy and accelerating particles~\citep{10.1103/RevModPhys.82.603, 10.1038/s42254-021-00419-x}---see Figure~\ref{fig10}. In~relativistic plasmas, this process can produce power-law particle distributions with spectral indices similar to those inferred from observations~\citep{10.1086/337972, 10.1086/171296}.

PIC simulations of relativistic reconnection show that the acceleration efficiency depends on several factors, including the magnetization parameter:
\begin{equation}
\sigma = \frac{B^2}{4\pi \rho c^2}
\end{equation}
where $B$ is the magnetic field strength and $\rho$ is the plasma density~\citep{10.1088/2041-8205/783/1/L21, 10.1088/0004-637X/806/2/167}. In~highly magnetized plasmas ($\sigma \gg 1$), reconnection can convert a significant fraction of the magnetic energy into particle kinetic energy~\citep{10.1007/s11214-014-0132-9}. The~resulting particle distributions typically exhibit power-law tails extending to the highest energies accessible in the \mbox{simulation~\citep{10.3847/2041-8205/816/1/L8, 10.1093/mnras/stw1620}.}

The maximum particle energy achievable through magnetic reconnection can be estimated as:
\begin{equation}
\gamma_{\max} \sim \frac{\sigma}{4} \frac{L}{r_L}
\end{equation}
where $L$ is the system size and $r_L = \gamma m c^2 / (eB)$ is the Larmor radius of the~particle.

The role of turbulence in particle acceleration has also been investigated through PIC simulations~\citep{10.1063/1.3703318, 10.1103/PhysRevLett.121.255101}. Relativistic turbulence can develop in jets through various instabilities, including the Kelvin-Helmholtz instability at the jet boundary and current-driven instabilities in the jet interior~\citep{10.1088/2041-8205/746/2/L14, 10.1103/PhysRevLett.111.015005}. Turbulent acceleration can produce extended particle distributions and may be particularly important in the jet spine where the plasma is highly relativistic~\citep{10.1093/mnras/sty854, 10.1103/PhysRevLett.118.055103}.

The turbulent acceleration rate can be approximated as:
\begin{equation}
\frac{d\gamma}{dt} \sim \frac{c}{L_{\textrm{turb}}} \left(\frac{\delta B}{B}\right)^2 \gamma
\end{equation}
where $L_{\textrm{turb}}$ is the turbulent correlation length and $\delta B/B$ is the relative magnetic field fluctuation~amplitude.

Shock acceleration remains another important mechanism for particle acceleration in jets~\citep{10.1086/590248, 10.1088/0004-637X/695/2/L189}. Relativistic shocks can form when fast-moving plasma encounters obstacles or when different regions of the jet have different velocities~\citep{10.1086/377260, 10.1088/0004-637X/726/2/75}. The~efficiency of shock acceleration depends on the shock parameters, including the Mach number and the magnetic field orientation~\citep{10.1088/0004-637X/783/2/91, 10.1103/PhysRevLett.114.085003}. Recent PIC simulations have shown that relativistic shocks can produce power-law particle distributions, but~the spectral index and maximum energy depend sensitively on the shock parameters~\citep{10.1088/0004-637X/726/2/75, 10.1103/PhysRevLett.113.155005}.

For a relativistic shock with Lorentz factor $\Gamma_s$, the~maximum particle energy is limited by synchrotron losses:
\begin{equation}
\gamma_{\max} \sim \sqrt{\frac{6\pi m c}{\sigma_T B}} \Gamma_s
\end{equation}
where $\sigma_T$ is the Thomson scattering~cross-section.

Coupling PIC results with GRMHD remains a challenge but is essential for connecting horizon-scale simulations with observable emission~\citep{10.3847/1538-4365/aab114, 10.1029/2018JA025713, 10.1007/s11214-025-01142-0}. The~computational requirements for fully kinetic simulations of jets are prohibitive, as~the relevant scales range from the plasma skin depth ($c/\omega_p$) to the system size ($\sim$ $10^6~r_g$)~\citep{10.1088/0004-637X/771/1/54}. Hybrid approaches that treat ions as particles and electrons as a fluid, or~that use reduced kinetic models, offer promising compromises between accuracy and computational efficiency~\citep{10.1086/513599, 10.1016/j.jcp.2013.11.035}.

Recent developments in hybrid PIC-MHD methods attempt to bridge this gap by using PIC simulations to calibrate subgrid models for particle acceleration in MHD simulations~\citep{10.1063/1.3703318, 10.3847/1538-4357/ab4c33}. These approaches use PIC simulations of local plasma processes to determine acceleration rates and particle distributions, which are then incorporated into global MHD simulations as source terms~\citep{10.1088/0004-637X/733/1/63, 10.1088/1475-7516/2011/05/026}. While promising, these methods require careful validation and may miss important non-local~effects.

The plasma composition of jets remains an open question with important implications for particle acceleration and emission~\citep{10.1038/26675, 10.1093/mnras/stab163}. Electron-positron plasmas have different acceleration properties compared to electron-proton plasmas, particularly in magnetic reconnection events~\citep{10.1002/2014GL062034, 10.1103/PhysRevLett.118.085101}. The~pair content of jets can be probed through polarization observations and the detection of gamma-ray absorption features, but~current observations provide only weak constraints~\citep{10.3390/galaxies6010005, 10.1088/0004-637X/804/2/111}.

\subsection{Radiative Feedback and~Observables}

\begin{figure}[H]
\includegraphics[width=115mm]{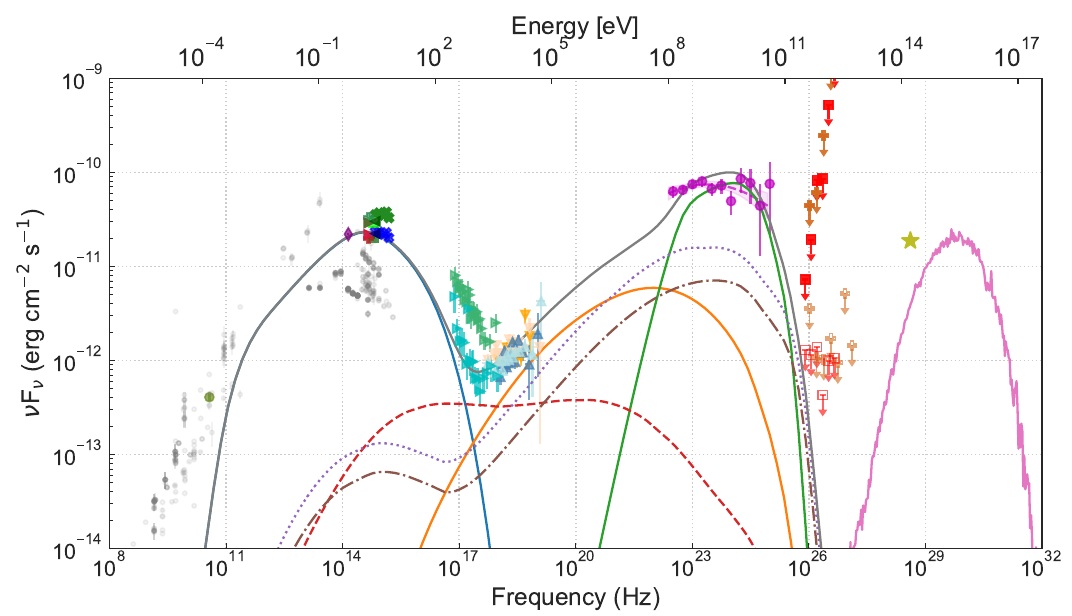}

\caption{\hl{Spectral Energy Distribution (SED)} 
 of PKS 0735+178 for December 2021, presented with a model that incorporates radiative contributions from electrons, protons, and~an external photon field. For a detailed description of each of the elements presented in the image, check the cited reference. (Credit: VERITAS and HESS collaborations~\citep{10.48550/arXiv.2306.17819}).}
\label{fig11}
\end{figure}

Incorporating radiation into simulations is another frontier that is essential for connecting theoretical models with observational data~\citep{10.3847/0004-637X/829/1/11, 10.3847/1538-4357/ab96c6}. Radiative transfer models that self-consistently evolve synchrotron, inverse Compton, and~hadronic processes are needed to connect theory with data from CTA, LSST, and~the EHT~\citep{10.1088/0004-637X/706/1/497, 10.1093/mnras/staa922}. This challenge is compounded by the multi-scale nature of jet emission, which spans from sub-horizon particle dynamics to kiloparsec-scale shocks~\citep{10.1088/0004-637X/768/1/54}. To~better understand these emission mechanisms, Figure~\ref{fig11} illustrates a typical blazar's Spectral Energy Distribution, clearly showing its synchrotron and inverse Compton~components.

The development of general relativistic radiative transfer codes has ushered in more realistic modeling of emission from the vicinity of black holes~\citep{10.1086/500349, 10.1088/0067-0049/184/2/387}. These codes solve the radiative transfer equation in curved spacetime, accounting for gravitational redshift, light bending, and~frame dragging effects~\citep{10.1088/0004-637X/701/2/1357, 10.1088/0004-637X/755/2/133}. The~resulting synthetic observations can be directly compared with EHT data, delimiting constraints on the physical conditions in the emission region~\citep{10.1093/mnras/stx587, 10.3847/1538-4357/aab6a8}.

Synchrotron emission from relativistic electrons in magnetic fields is the dominant radiation mechanism at radio through 
optical wavelengths~\citep{1986rpa..book.....R}. The~synchrotron power radiated by an electron with Lorentz factor $\gamma$ in a magnetic field $B$ is:
\begin{equation}
P_{\textrm{sync}} = \frac{4}{3} \sigma_T c \beta^2 \gamma^2 \frac{B^2}{8\pi}
\end{equation}
where $\beta \approx 1$ for relativistic~particles.

The emission properties depend on the electron distribution function, magnetic field strength, and~geometry~\citep{10.1086/152724, 10.1086/163592}. GRMHD simulations can predict the magnetic field structure and bulk plasma properties, but~the electron distribution must be specified through additional assumptions or microphysical modeling~\citep{10.3847/2041-8213/ab9532, 10.3847/0004-637X/829/1/11}.

Inverse Compton scattering of low-energy photons by relativistic electrons produces high-energy emission in jets~\citep{10.1086/173251, 10.1086/173633}. The~power radiated through inverse Compton scattering is:
\begin{equation}
P_{\text{IC}} = \frac{4}{3} \sigma_T c \beta^2 \gamma^2 U_{\text{ph}}
\end{equation}
where $U_{\text{ph}}$ is the energy density of the seed photon~field.

The seed photons can originate from synchrotron emission within the jet (synchrotron self-Compton, SSC), external radiation fields from the accretion disk or broad line region (external Compton, EC), or~the cosmic microwave background~\citep{10.1111/j.1365-2966.2009.15898.x, 10.3847/0004-637X/830/2/94}. The~relative importance of these different seed photon populations depends on the location of the emission region and the external photon field density~\citep{10.1111/j.1745-3933.2010.00884.x, 10.3847/0004-637X/826/2/115}.

The ratio of inverse Compton to synchrotron luminosity, known as the Compton dominance parameter, is given by:
\begin{equation}
q = \frac{L_{\text{IC}}}{L_{\text{sync}}} = \frac{U_{\text{ph}}}{U_B}
\end{equation}
where $U_B = B^2/(8\pi)$ is the magnetic energy~density.

Hadronic processes, including proton-proton collisions and photo-meson production, can contribute to high-energy emission and produce neutrinos and cosmic rays~\mbox{\citep{1993A&A...269...67M, 10.1093/mnras/stac3190}}. These processes require the acceleration of protons to ultra-relativistic energies and sufficient target material or photon fields~\citep{10.1086/346261, 10.1088/0004-637X/768/1/54}. The~efficiency of hadronic acceleration and the resulting emission depend on the plasma composition and the magnetic field structure~\citep{10.1093/mnras/stv179, 10.1093/mnras/stz2380}.

The threshold energy for photo-meson production is:
\begin{equation}
E_{p,\text{th}} = \frac{m_\pi c^2 (m_\pi + 2m_p) c^2}{4 E_{\text{ph}}} \approx \frac{7 \times 10^{16} \text{ eV}}{E_{\text{ph}} / \text{eV}}
\end{equation}
where $m_\pi$ and $m_p$ are the pion and proton masses, respectively, and~$E_{\text{ph}}$ is the target photon~energy.

Radiative cooling can significantly affect the dynamics of jets, particularly in luminous sources where the cooling timescale becomes comparable to the dynamical timescale~\mbox{\citep{10.1088/0004-637X/763/2/134, 10.1111/j.1365-2966.2007.12758.x}}. The~synchrotron cooling timescale is:
\begin{equation}
t_{\text{sync}} = \frac{6\pi m_e c}{\sigma_T B^2 \gamma} \approx 1.3 \times 10^9 \left(\frac{B}{\text{G}}\right)^{-2} \left(\frac{\gamma}{10^4}\right)^{-1} \text{ s}
\end{equation}

Synchrotron and inverse Compton cooling can reduce the electron energy and modify the emission properties~\citep{10.1111/j.1365-2966.2005.09494.x, 10.1111/j.1365-2966.2010.18140.x}. In~extreme cases, radiative cooling can affect the jet dynamics through radiation pressure and energy loss~\citep{10.1103/PhysRevD.111.083049, 10.3390/galaxies4030015, 10.3847/0004-637X/832/1/12}.

The implementation of radiative transfer in GRMHD simulations remains computationally challenging due to the need to solve the radiative transfer equation at each grid point and time step~\citep{10.1093/mnras/stt1881, 10.3847/1538-4357/ab29ff}. Approximate methods, such as the M1 closure or flux-limited diffusion, are often used to reduce the computational cost~\citep{10.1016/0022-4073(84)90112-2, 10.1093/mnras/stw3116}. However, these approximations may not be accurate in optically thin regions or when the radiation field is highly anisotropic~\citep{10.1088/0067-0049/199/1/14, 10.1093/mnras/stv2854}.

Recent advances in GPU computing and machine learning techniques offer new possibilities for accelerating radiative transfer calculations~\citep{10.1186/s40668-017-0020-2, 10.1103/RevModPhys.91.045002}. Neural networks can be trained to approximate the radiative transfer solution, potentially reducing the computational cost by orders of magnitude~\citep{10.1073/pnas.191278911, 10.48550/arXiv.1902.10159}. These methods are still in development but show promise for enabling self-consistent radiative GRMHD~simulations.

\subsection{Jet-Environment~Interactions}

The interaction between jets and their surrounding environment  plays an important role in determining jet morphology and the efficiency of AGN feedback~\citep{10.1016/j.newar.2020.101539, 10.1093/mnras/sty067}. Jets must propagate through the interstellar medium of the host galaxy and potentially through the intracluster medium in galaxy clusters~\citep{10.1146/annurev.astro.45.051806.110625, 10.1146/annurev-astro-081811-125521}. These interactions involve complex multi-phase physics that is challenging to model numerically~\citep{10.1111/j.1365-2966.2012.21479.x, 10.1093/mnras/stw1368}.

\begin{figure}[H]
\includegraphics[width=115mm]{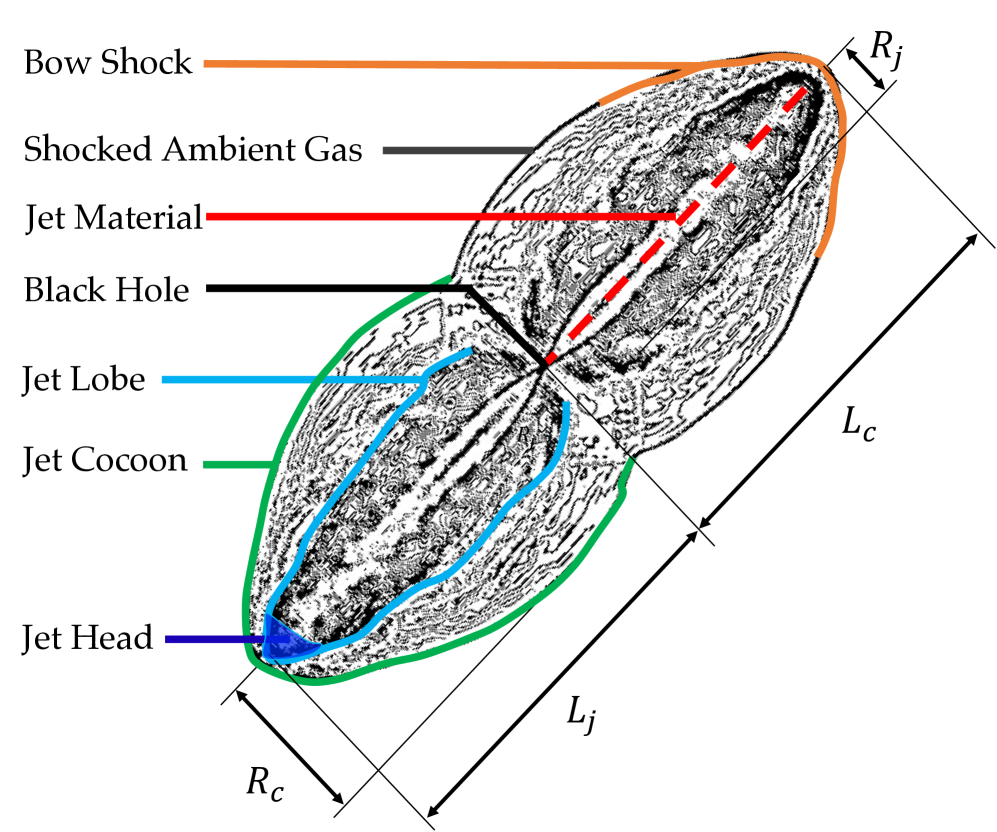}

\caption{Jet structure derived from a fiducial simulation slice plot. It features a central jet (length $L_J$, radius $R_J$) terminating at a jet head, which inflates a lobe of shocked jet material. This lobe is enveloped by a larger cocoon of shocked ambient gas, with~a bow shock defining the overall cocoon length ($L_c$) and radius ($R_c$) (Credit: GADGET4-OSAKA simulations~\citep{10.48550/arXiv.2508.21282}).}
\label{fig12}
\end{figure}

Hydrodynamic simulations of jet propagation have revealed several key insights into jet-environment interactions~\citep{1982A&A...113..285N, 10.1086/184799}. Jets can efficiently entrain ambient material, leading to the formation of cocoons and backflows that can affect the host galaxy~\citep{10.1103/RevModPhys.56.255, 10.48550/arXiv.astro-ph/0007261}. The~entrainment rate depends on the jet parameters, including the power, opening angle, and~density, as~well as the properties of the ambient medium~\citep{10.1093/mnras/stw2944, 10.1051/0004-6361:20021649}. These complex interactions, including the formation of characteristic cocoons and lobes, are  illustrated in Figure~\ref{fig12}.
\vspace{-6pt}

The mass entrainment rate can be estimated as:
\begin{equation}
\dot{M}_{\text{ent}} \sim \rho_{\text{amb}} v_j A_j \left(\frac{\delta v}{v_j}\right)
\end{equation}
where $\rho_{\text{amb}}$ is the ambient density, $v_j$ is the jet velocity, $A_j$ is the jet cross-sectional area, and~$\delta v$ is the velocity difference at the jet~boundary.

The Kelvin-Helmholtz instability at the jet boundary can lead to mixing between the jet and ambient material~\citep{1991bja..book..278B, 10.1051/0004-6361/200913012}. This mixing can reduce the jet velocity and increase the mass loading, affecting the jet propagation and termination~\citep{10.1051/0004-6361:200809687, 10.1088/0004-637X/705/2/1594}. The~growth rate of the instability depends on the velocity shear and density contrast between the jet and ambient medium~\citep{10.1086/308656, 10.1088/0004-637X/700/1/684}.

The Kelvin-Helmholtz growth rate is approximately:
\begin{equation}
\omega_{\text{KH}} \sim \frac{k \Delta v}{\sqrt{1 + \rho_j/\rho_{\text{amb}}}}
\end{equation}
where $k$ is the wavenumber, $\Delta v$ is the velocity difference, and~$\rho_j$ and $\rho_{\text{amb}}$ are the jet and ambient~densities.

Magnetic fields can stabilize jets against disruption by hydrodynamic instabilities~\mbox{\citep{2000A&A...355..818A, 10.1093/mnras/stu196}}. Magnetized jets can maintain their collimation over larger distances and are more efficient at transporting energy to large scales~\citep{10.1086/501499, 10.1111/j.1365-2966.2012.20721.x}. However, the~magnetic field configuration can also affect the interaction with the ambient medium, with~helical fields potentially enhancing entrainment through magnetic reconnection~\citep{10.1086/510361, 10.3847/0004-637X/824/1/48}.

The formation of shocks in the ambient medium can lead to heating and acceleration of the surrounding gas~\citep{10.1111/j.1365-2966.2009.14811.x, 10.1088/0004-637X/757/2/136}. These shocks can drive outflows and turbulence in the host galaxy, potentially affecting star formation and gas accretion onto the central black hole~\citep{1998A&A...331L...1S, 10.1086/379143}. The~efficiency of this feedback depends on the jet power and the properties of the ambient medium~\citep{10.1111/j.1365-2966.2005.09675.x, 10.1111/j.1365-2966.2007.12153.x}.

The shock heating rate can be expressed as:
\begin{equation}
\dot{E}_{\text{heat}} \sim \frac{P_{\text{jet}}}{V_{\text{cocoon}}} \times f_{\text{coupling}}
\end{equation}
where $V_{\text{cocoon}}$ is the volume of the shocked region and $f_{\text{coupling}}$ is the coupling efficiency between jet and ambient~medium.

Recent simulations have begun to include more realistic prescriptions for the ambient medium, including multi-phase gas with different temperatures and densities~\citep{10.1111/j.1365-2966.2012.21479.x, 10.1093/mnras/sty067}. These simulations show that jets can have different effects on different gas phases, potentially leading to complex feedback processes~\citep{10.1093/mnras/stx2269, 10.1093/mnras/stw2944}. Cold gas clouds can be disrupted or accelerated by jets, while hot gas can be heated and driven out of the galaxy~\citep{10.1086/425494, 10.1093/mnras/stw1368}.

The role of cosmic rays accelerated in jets is another important consideration for jet-environment interactions~\citep{10.3847/1538-4357/834/2/208, 10.1093/mnras/stw2941}. Cosmic rays can diffuse away from jets and deposit energy in the ambient medium through various processes, including streaming instabilities and hadronic collisions~\citep{10.1111/j.1365-2966.2012.21533.x, 10.1111/j.1365-2966.2005.09953.x}. This energy deposition can introduce a distributed heating source that affects the thermal balance of the ambient medium~\citep{10.1111/j.1365-2966.2007.12692.x, 10.1088/2041-8205/797/2/L18}.

The cosmic ray diffusion coefficient in the ambient medium is typically:
\begin{equation}
D_{\text{CR}} \sim \frac{1}{3} c \lambda_{\text{mfp}} \sim \frac{c r_L}{3} \left(\frac{B}{\delta B}\right)^2
\end{equation}
where $\lambda_{\text{mfp}}$ is the mean free path and $\delta B/B$ is the magnetic field fluctuation~level.

\section{Open Questions and Future~Directions}\label{sec5}

Despite rapid progress in both observational and theoretical studies of black hole jets, several fundamental questions remain unresolved. These open questions span multiple scales and physical regimes, from~quantum processes near the event horizon to the large-scale impact of jets on galaxy evolution. Addressing these questions will require coordinated efforts combining next-generation observations, advanced theoretical modeling, and~innovative computational~techniques.

\subsection{Fundamental Physics~Questions}

\subsubsection{Jet Composition and Plasma~Physics}
One of the most fundamental open questions concerns the composition of relativistic jets~\citep{10.1038/26675, 10.1093/mnras/stab163}. Are jets primarily electron-proton or electron-positron dominated? This question has profound implications for jet energetics, particle acceleration mechanisms, and~observational signatures. Electron-positron jets would be lighter and potentially more efficient at converting magnetic energy into kinetic energy, while electron-proton jets would carry more mass and could contribute significantly to cosmic ray production~\citep{10.1086/429314, 10.1088/0004-637X/768/1/54}.

The pair-to-proton ratio in jets can be parameterized as:
\begin{equation}
  \frac{n_{e^+}}{n_p}
\end{equation}
where $n_{e^+}$ and $n_p$ are the positron and proton number densities. Current observational constraints suggest $ \sim $1--20, but~with large uncertainties~\citep{10.3390/galaxies6010005}.

Current observational evidence on jet composition come primarily from polarization measurements and spectral modeling~\citep{10.3390/galaxies6010005, 10.1088/0004-637X/804/2/111}. The~degree of linear polarization can trace information about the magnetic field structure and particle distribution, while circular polarization can probe Faraday rotation effects that depend on the plasma \mbox{composition~\citep{10.1086/429157, 10.1086/504715}}. However, these estimates are often model-dependent and subject to degeneracies between different physical parameters~\citep{10.1111/j.1365-2966.2005.08954.x, 10.1088/0004-637X/725/1/750}.

Future polarization observations with the EHT and other facilities may help resolve this question by furnishing detailed maps of the magnetic field structure and Faraday rotation in jet launching regions~\citep{10.3847/2041-8213/abe71d, 10.1142/S0218271817300014}. The~detection of gamma-ray absorption features in blazar spectra could also yield limits on the pair content of jets~\citep{10.1103/PhysRev.155.1404, 10.1016/S0927-6505(02)00155-X}. Additionally, neutrino observations may help distinguish between leptonic and hadronic jet models, as~only hadronic processes can produce high-energy neutrinos~\citep{10.1093/mnras/stac3190, 10.1093/mnras/sty1852}.

The question of jet composition is closely related to the broader issue of plasma physics in relativistic outflows~\citep{10.1088/0004-637X/771/1/54, 10.1093/mnras/stx2530}. How do kinetic processes such as magnetic reconnection and turbulence affect the bulk properties of jets? What role do plasma instabilities play in jet collimation and particle acceleration? These questions require detailed kinetic simulations that can capture the full complexity of relativistic plasma physics~\citep{10.1103/PhysRevLett.106.195003, 10.1103/PhysRevLett.118.055103}.

\subsubsection{Acceleration Sites and~Mechanisms}
What is the dominant mechanism—shocks, turbulence, or~reconnection—for accelerating particles to ultra-relativistic energies? This question is central to understanding the high-energy emission from jets and their role as cosmic ray accelerators~\citep{10.1088/2041-8205/783/1/L21, 10.1093/mnras/stu2364}. Different acceleration mechanisms predict different particle distributions, variability timescales, and~emission signatures~\citep{10.1086/309533, 10.1093/mnras/stt1494}.

Shock acceleration has been the traditional paradigm for particle acceleration in jets, based on the success of diffusive shock acceleration theory in explaining cosmic ray spectra~\citep{10.1093/mnras/182.2.147, 10.1086/182658}. However, the~rapid variability observed in some blazars challenges this paradigm, as~it implies acceleration timescales much shorter than those expected from shock acceleration~\citep{10.1086/520635, 10.1086/521382}. This has led to increased interest in alternative mechanisms such as magnetic reconnection and turbulent acceleration~\citep{10.1111/j.1745-3933.2009.00635.x, 10.1088/2041-8205/783/1/L21}.

The acceleration timescale for different mechanisms can be compared:
\begin{align}
t_{\text{shock}} &\sim \frac{r_L}{c} \left(\frac{c}{v_s}\right)^2 \
t_{\text{reconnection}} &\sim \frac{L}{c} \frac{1}{\sigma} \
t_{\text{turbulence}} &\sim \frac{L_{\text{turb}}}{c} \left(\frac{B}{\delta B}\right)^2
\end{align}
where $v_s$ is the shock velocity, $L$ is the system size, and~$L_{\text{turb}}$ is the turbulent correlation~length.

Magnetic reconnection can deliver rapid acceleration through the direct conversion of magnetic energy into particle kinetic energy~\citep{10.1086/337972, 10.1088/0004-637X/771/1/54}. Recent PIC simulations have shown that relativistic reconnection can produce power-law particle distributions with spectral indices consistent with observations~\citep{10.1103/PhysRevLett.113.155005, 10.3847/2041-8205/816/1/L8}. However, the~efficiency of reconnection in realistic jet environments and its ability to accelerate particles to the highest observed energies remain uncertain~\citep{10.1103/PhysRevLett.105.235002, 10.1088/0004-637X/782/2/104}.

Turbulent acceleration through stochastic interactions with magnetic fluctuations offers another possibility for rapid particle acceleration~\citep{10.1007/978-3-662-04814-6, 10.1093/mnras/sty854}. This mechanism can operate continuously throughout the jet and may be particularly important in regions where the plasma is highly magnetized~\citep{10.1063/1.3703318, 10.1103/PhysRevLett.121.255101}. However, the~development and maintenance of turbulence in relativistic jets is not well understood, and~the efficiency of turbulent acceleration depends on the turbulence spectrum and magnetic field structure~\citep{10.1103/PhysRevD.99.083006, 10.1103/PhysRevLett.118.245101}. Sebastian and Comisso~\citep{10.3847/2041-8213/ae1696} and  Das~et~al.~\citep{10.48550/arXiv.2506.04212} apply 3D Particle-in-Cell (PIC) simulations to investigate how turbulence in strongly magnetized (high-$\sigma$) astrophysical plasmas accelerates particles to non-thermal, high energies. They explore different, recently proposed mechanisms beyond traditional scattering-based theories and highlight complementary mechanisms within the same environment: (i) Mirror acceleration that relies on compressive modes (magnetic pressure); and (ii) Curvature-drift acceleration that relies on shear/torsional modes (magnetic tension).

Future observations with CTA and other high-energy facilities will differentiate between acceleration mechanisms through detailed studies of variability timescales and spectral evolution during flares~\citep{10.1142/10986}. The~correlation of gamma-ray variability with emission at other wavelengths can help identify the location and mechanism of particle acceleration~\citep{10.1088/2041-8205/710/2/L126, 10.1088/0004-637X/715/1/362}. Multi-wavelength campaigns coordinated with EHT observations may directly connect horizon-scale magnetic field structures with particle acceleration processes~\citep{10.3847/1538-4357/abac0d}.

\subsection{Jet Dynamics and~Structure}
\unskip

\subsubsection{Jet Stability and~Collimation}
How do magnetic instabilities, entrainment, and~interaction with the external medium determine jet morphology? Understanding jet stability is necessary for explaining the observed diversity of jet morphologies, from~highly collimated pencil beams to wide-angle outflows~\citep{10.1016/j.newar.2020.101539, 10.3847/1538-3881/ac4d23}. The~stability of jets depends on the balance between magnetic pressure, gas pressure, and~external confinement~\citep{10.1103/RevModPhys.56.255, 10.1111/j.1365-2966.2007.12050.x}.

The jet opening angle can be estimated from the balance of forces:
\begin{equation}
\tan \theta_j \sim \sqrt{\frac{P_{\text{gas}} + P_{\text{ext}}}{P_{\text{mag}}}}
\end{equation}
where $P_{\text{gas}}$, $P_{\text{ext}}$, and~$P_{\text{mag}}$ are the gas, external, and~magnetic pressures, respectively.

Magnetic instabilities such as the current-driven kink mode can disrupt jet collimation and lead to helical distortions~\citep{10.1093/mnras/stu196, 10.3847/0004-637X/824/1/48}. The~growth rate of these instabilities depends on the magnetic field configuration and the plasma beta parameter~\citep{2000A&A...355..818A, 10.1111/j.1365-2966.2012.20721.x}. While some instabilities can be disruptive, others may play a beneficial role in jet collimation through magnetic pinching effects~\citep{10.1093/mnras/279.2.389, 10.1007/978-3-540-76937-8_9}.

The kink instability growth rate is approximately:
\begin{equation}
\omega_{\text{kink}} \sim \frac{v_A}{R_j} \sqrt{\frac{k_z R_j}{1 + k_z R_j}}
\end{equation}
where $v_A$ is the Alfvén velocity, $R_j$ is the jet radius, and~$k_z$ is the axial~wavenumber.

Entrainment of ambient material can also affect jet stability and propagation~\citep{10.1093/mnras/stw2944, 10.1051/0004-6361:20021649}. The~entrainment rate depends on the velocity shear at the jet boundary and the density contrast between the jet and ambient medium~\citep{10.1051/0004-6361/200913012, 10.1088/0004-637X/705/2/1594}. Excessive entrainment can lead to jet deceleration and eventual termination, while moderate entrainment may help stabilize the jet through mass loading~\citep{10.1051/0004-6361:200809687, 10.1093/mnras/stw1368}.

The external pressure profile also plays a fundamental role in jet collimation~\citep{10.48550/arXiv.astro-ph/0007261, 10.1111/j.1365-2966.2009.14887.x}. Jets propagating through declining pressure profiles will naturally expand and decelerate, while those in increasing pressure environments may be further collimated~\citep{10.1086/184799, 10.1088/0004-637X/757/2/136}. The~pressure profile depends on the host galaxy properties and can vary significantly between different environments~\citep{10.1111/j.1365-2966.2012.21479.x, 10.1093/mnras/stw2944}.

Future numerical simulations with improved resolution and physics will help address these questions by following jet evolution over larger spatial and temporal scales~\citep{10.3847/0067-0049/225/2/22, 10.3847/1538-4365/ac9966}. The~inclusion of realistic ambient medium properties and magnetic field configurations will be crucial for understanding jet stability in different environments~\citep{10.1093/mnras/sty067, 10.1093/mnras/stw2941}.

\subsubsection{Energy Dissipation and~Transport}
How is energy transported and dissipated along jets from the launching region to the terminal shocks? This question is fundamental to understanding jet energetics and the efficiency of AGN feedback~\citep{10.1146/annurev-astro-081811-125521, 10.1146/annurev.astro.45.051806.110625}. Energy can be transported in various forms, including kinetic energy of the bulk flow, magnetic energy in ordered and turbulent fields, and~the energy of relativistic particles~\citep{10.1111/j.1365-2966.2007.12050.x, 10.1111/j.1365-2966.2012.22002.x}.

The energy flux in different forms can be written as:
\begin{align}
\dot{E}_{\text{kinetic}} &= \frac{1}{2} \dot{M} v^2 \
\dot{E}_{\text{magnetic}} &= \frac{B^2}{4\pi} A v \
\dot{E}_{\text{Poynting}} &= \frac{c}{4\pi} \mathbf{E} \times \mathbf{B} \cdot \hat{\mathbf{n}} A
\end{align}
where $A$ is the jet cross-sectional area and $\hat{\mathbf{n}}$ is the normal~vector.

The conversion between different energy forms occurs through various dissipation mechanisms~\citep{10.1088/2041-8205/783/1/L21, 10.1088/0004-637X/809/1/38}. Magnetic reconnection can convert magnetic energy into particle kinetic energy and heat~\citep{10.1088/0004-637X/771/1/54, 10.1093/mnras/stx2530}. Shocks can thermalize kinetic energy and accelerate particles~\citep{10.1086/590248, 10.1088/0004-637X/783/2/91}. Turbulence can cascade energy from large to small scales where it can be dissipated through various microphysical processes~\citep{10.1063/1.3703318, 10.1103/PhysRevLett.121.255101}.

The efficiency of energy transport depends on the jet structure and the presence of instabilities or other dissipation mechanisms~\citep{10.1111/j.1365-2966.2006.10256.x, 10.1088/0004-637X/711/1/50}. Highly magnetized jets may be more efficient at transporting energy to large scales, while kinetically dominated jets may dissipate energy more rapidly through internal shocks and turbulence~\citep{10.1088/0004-637X/698/2/1570, 10.1088/0004-637X/696/2/1142}.

Observational signatures of energy dissipation come from multi-wavelength studies of jet emission and morphology~\citep{10.1111/j.1365-2966.2009.14887.x, 10.1111/j.1365-2966.2006.10525.x}. The~radio luminosity of jets quantifies a measure of the energy dissipated through synchrotron emission, while X-ray and gamma-ray observations can probe higher-energy dissipation processes~\citep{10.1086/321394, 10.1126/science.1134408}. The~spatial distribution of emission along jets can also provide information about where energy dissipation occurs~\citep{10.1088/0004-637X/746/2/151, 10.1088/2041-8205/723/2/L207}.

\subsection{Feedback and Environmental~Impact}
\unskip

\subsubsection{Feedback~Efficiency}
How efficiently do jets deposit energy into galactic and cluster environments, and~how does this regulate galaxy evolution? This question is central to understanding the role of AGN feedback in galaxy formation and evolution~\citep{1998A&A...331L...1S, 10.1086/379143}. Jets can affect their environment through various mechanisms, including direct heating, driving of outflows, and~disruption of gas accretion~\citep{10.1111/j.1365-2966.2005.09675.x, 10.1111/j.1365-2966.2007.12153.x}.

The feedback efficiency can be quantified as:
\begin{equation}
\eta_{\text{feedback}} = \frac{\dot{E}_{\text{deposited}}}{\dot{E}_{\text{jet}}}
\end{equation}
where $\dot{E}_{\text{deposited}}$ is the rate of energy deposition in the ambient medium and $\dot{E}_{\text{jet}}$ is the jet~power.

The efficiency of jet feedback depends on the coupling between jets and the ambient medium~\citep{10.1016/j.newar.2020.101539, 10.1093/mnras/sty067}. This coupling is determined by factors such as the jet power, opening angle, and~duty cycle, as~well as the properties of the ambient medium including density, temperature, and~magnetic field strength~\citep{10.1111/j.1365-2966.2012.21479.x, 10.1093/mnras/stx2269}. The~feedback efficiency may also depend on the host galaxy properties and environment~\citep{10.1111/j.1365-2966.2012.20414.x, 10.1051/0004-6361/201833883}.

Observational evidence for jet feedback comes from studies of galaxy clusters, where jets from central AGN appear to balance radiative cooling of the intracluster medium~\citep{10.1146/annurev.astro.45.051806.110625, 10.1146/annurev-astro-081811-125521}. However, the~detailed mechanisms responsible for this balance are not well understood~\citep{10.3847/1538-4357/834/2/208}. How do jets maintain the observed temperature and entropy profiles in cluster cores? What role do sound waves, turbulence, and~cosmic rays play in distributing energy throughout the cluster?~\citep{10.1093/mnras/stw2941, 10.1088/2041-8205/797/2/L18}.

The cooling-heating balance in clusters can be expressed as:
\begin{equation}
\dot{E}_{\text{cooling}} = \Lambda(T) n_e n_H V \approx \dot{E}_{\text{jet}} \times \eta_{\text{feedback}}
\end{equation}
where $\Lambda(T)$ is the cooling function and $n_e$, $n_H$ are the electron and hydrogen~densities.

Future observations with LSST and other facilities will define statistical bounds on feedback efficiency through studies of large samples of galaxies and clusters~\citep{10.48550/arXiv.0912.0201, 10.3847/1538-4357/ab042c}. The~correlation between jet activity and host galaxy properties can reveal the conditions under which feedback is most effective~\citep{10.1088/0004-637X/782/1/9, 10.1038/s41550-018-0403-6}. Multi-wavelength observations can trace the impact of jets on different gas phases and provide insights into the physical mechanisms responsible for feedback~\citep{10.1088/0004-637X/757/2/136, 10.1093/mnras/stw2944}.

\subsubsection{Duty Cycles and~Triggering}
What determines the duty cycle of jet activity, and~how does it relate to black hole accretion and galaxy mergers? Understanding jet triggering is required for predicting when and where jets will be active and for assessing their cumulative impact on galaxy evolution~\citep{10.1093/mnras/stu327, 10.1088/0004-637X/701/1/66}.

The duty cycle can be defined as:
\begin{equation}
f_{\text{duty}} = \frac{t_{\text{active}}}{t_{\text{active}} + t_{\text{quiescent}}}
\end{equation}
where $t_{\text{active}}$ and $t_{\text{quiescent}}$ are the durations of active and quiescent~phases.

Current models suggest that jet activity may be episodic, with~periods of high activity alternating with quiescent phases~\citep{10.1146/annurev-astro-112420-035022, 10.1093/mnras/staa476}. The~duty cycle may depend on factors such as the black hole spin, accretion rate, and~magnetic field configuration~\citep{10.1088/0004-637X/711/1/50, 10.1093/mnras/stac285}. Galaxy mergers and interactions may also trigger jet activity by driving gas toward the central black hole and enhancing the magnetic flux~\citep{10.1086/499298, 10.1086/344675}.

Observational constraints on duty cycles come from studies of jet morphology and the statistics of AGN activity~\citep{10.1111/j.1365-2966.2012.20414.x, 10.1051/0004-6361/201833883}. Double-lobed radio sources suggest episodic activity, while the fraction of galaxies hosting active jets allows us to infer the average duty cycle \citep{10.1016/j.newar.2020.101539}. However, these inferences are limited by selection effects and the difficulty of measuring jet ages~\citep{10.1093/mnras/stx959}.

The jet age can be estimated from the advance speed:
\begin{equation}
t_{\text{jet}} \sim \frac{D_{\text{lobe}}}{v_{\text{advance}}} \sim \frac{D_{\text{lobe}}}{c} \left(\frac{\rho_{\text{jet}}}{\rho_{\text{amb}}}\right)^{1/3}
\end{equation}
where $D_{\text{lobe}}$ is the lobe distance and $v_{\text{advance}}$ is the advance~velocity.

Future time-domain surveys with LSST will produce unprecedented insights into jet duty cycles through long-term monitoring of AGN variability~\citep{10.1088/0004-637X/753/2/106, 10.3847/1538-4357/834/2/111}. The~survey will detect transitions between active and quiescent states and correlate these with host galaxy properties and environmental factors~\citep{10.1093/mnras/stz3244}. The~combination of optical monitoring with radio and X-ray observations will assemble a comprehensive view of jet activity cycles~\citep{10.48550/arXiv.1811.06542}.

\subsection{Multi-Messenger~Connections}
\unskip

\subsubsection{Cosmic Ray and Neutrino~Production}
What fraction of high-energy cosmic rays and neutrinos originate in black hole jets? This question connects jet physics to fundamental questions about the origin of cosmic rays and the sources of high-energy neutrinos~\citep{10.1093/mnras/stac3190, 10.1093/mnras/sty1852}. Jets are among the few astrophysical sources capable of accelerating particles to the highest observed energies~\citep{10.1088/1475-7516/2010/10/013, 10.3389/fspas.2019.00023}.

The production of ultra-high-energy cosmic rays (UHECRs) requires acceleration to energies exceeding $E > 10^{20}$ eV~\citep{10.1088/1475-7516/2010/10/013}. This requires either extremely powerful acceleration mechanisms or very large acceleration regions~\citep{10.1093/mnras/stt179}. Jets from AGN are promising candidates for UHECR acceleration, but~the acceleration efficiency and maximum energies achievable in realistic jet environments remain uncertain~\citep{10.1088/0004-637X/749/1/63, 10.3847/1538-4357/aaa7ee}.

The maximum energy for cosmic ray acceleration in jets is limited by various factors:
\begin{align}
E_{\max,\text{sync}} &\sim \sqrt{\frac{6\pi m_p c^3}{\sigma_T B}} \approx 10^{20} \left(\frac{B}{\text{G}}\right)^{-1/2} \text{ eV} \\
E_{\max,\text{IC}} &\sim \sqrt{\frac{6\pi m_p c^5}{\sigma_T U_{\text{ph}}}} \\
E_{\max,\text{escape}} &\sim \frac{eB R}{20} \approx 10^{20} \left(\frac{B}{\text{G}}\right) \left(\frac{R}{\text{pc}}\right) \text{ eV}
\end{align}

Neutrino production in jets requires hadronic processes such as proton-proton collisions or photo-meson production~\citep{1993A&A...269...67M, 10.1093/mnras/stac3190}. The~efficiency of these processes depends on the proton content of jets and the availability of target material or photon fields~\citep{10.1093/mnras/stv179, 10.1093/mnras/stz2380}. Current limits from IceCube are beginning to narrow down hadronic models for blazar emission, but~definitive detection of neutrinos from jets remains elusive~\citep{10.1126/science.aat2890, 10.1088/1748-0221/12/03/P03012}.

The neutrino production rate through photo-meson interactions is:
\begin{equation}
\dot{N}_\nu \sim f_\pi \times n_p \times \sigma_{p\gamma} \times c \times n_\gamma \times V
\end{equation}
where $f_\pi \approx 0.2$ is the fraction of proton energy transferred to pions, $\sigma_{p\gamma}$ is the photo-meson cross-section, and~$n_p$, $n_\gamma$ are the proton and photon~densities.

Future observations with next-generation neutrino detectors such as IceCube-Gen2 and KM3NeT will provide improved sensitivity to neutrino emission from jets~\citep{10.1088/1361-6471/abbd48}, however recent observations may already represent landmark discoveries in multi-messenger astrophysics, connecting cosmic-ray acceleration with extreme phenomena in the universe. These observations are: (i) the brightest gamma-ray burst (GRB) ever recorded, GRB 221009A~\citep{10.1126/science.adg9328, 10.3847/2041-8213/acc39c}; (ii) the highest-energy neutrino event KM3-230213A, from~NGC 1068  (a Seyfert galaxy that shows a steady neutrino flux from a decade of data) \cite{10.1126/science.abg3395}; and (iii)~the Telescope Array observation of the Amaterasu particle, an~ultra-high-energy cosmic ray with an estimated energy of $\sim$244 EeV (2.44 $\times$ 10$^{20}$ eV)~\citep{10.1126/science.abo5095} that may have originated from NGC 1068 – the same nearby Seyfert galaxy identified by IceCube as a source of high-energy neutrinos, although~\citep{10.3847/2041-8213/ade99f} proposes the blazar PKS 1717+177 as a candidate source for the Amaterasu~particle.

Furthermore, the~correlation of neutrino detections with gamma-ray flares observed by CTA can help distinguish between leptonic and hadronic emission models~\citep{10.1142/10986}. The~detection of neutrinos from individual sources would constitute definitive evidence for hadronic acceleration in jets~\citep{10.3847/1538-4357/aaa7ee}. UHECRs remain one of the most intriguing puzzles in astroparticle physics. While recent data have provided strong hints about their origins—particularly pointing to nearby starburst galaxies—definitive identification of sources and acceleration mechanisms will require more data, better composition measurements, and~deeper multi-messenger correlations~\citep{10.1016/j.physrep.2019.01.002}.

\subsubsection{Gravitational Wave~Associations}
How do jets relate to gravitational wave sources, particularly black hole mergers and neutron star disruptions? The detection of gravitational waves has opened a new window into extreme astrophysical phenomena and may unveil insights into jet formation in previously unexplored regimes~\citep{10.1103/PhysRevLett.116.061102, 10.3847/2041-8205/821/1/L18}.

Black hole mergers detected by LIGO/Virgo may be associated with electromagnetic counterparts powered by jets, particularly in asymmetric mergers or mergers involving neutron stars~\citep{10.1111/j.1365-2966.2010.16864.x, 10.1093/mnrasl/slx175}. The~efficiency of jet launching in merger events depends on the magnetic field configuration and the properties of any surrounding material~\citep{10.3847/2041-8205/824/1/L6, 10.1103/PhysRevD.95.063016}. Numerical relativity simulations are beginning to explore these scenarios, but~observational data remains limited~\citep{10.1103/PhysRevLett.111.061105}.

The jet power from black hole mergers can be estimated as:
\begin{equation}
P_{\text{jet}} \sim \frac{a_*^2 \Phi^2 c}{6\pi M^2} \sim 10^{52} \left(\frac{a_*}{0.9}\right)^2 \left(\frac{\Phi}{10^{30} \text{ G cm}^2}\right)^2 \left(\frac{M}{60 M_\odot}\right)^{-2} \text{ erg s}^{-1}
\end{equation}

Neutron star mergers, such as GW170817, can produce jets that power gamma-ray bursts and kilonovae~\citep{10.1103/PhysRevLett.119.161101, 10.3847/2041-8213/aa8f41}. The~properties of these jets and their interaction with merger ejecta offer insights into jet formation in neutron star environments~\citep{10.1038/s41586-018-0486-3, 10.1126/science.aau8815}. The~detection of gravitational waves from neutron star mergers triggers advance warning for electromagnetic follow-up observations~\citep{10.1088/978-0-7503-1369-8}.

Future gravitational wave detectors with improved sensitivity will detect more merger events and potentially reveal new classes of sources~\citep{10.48550/arXiv.1907.04833}. Space-based detectors such as LISA will extend the sensitivity to lower frequencies, potentially detecting the inspiral phase of massive black hole mergers~\citep{10.48550/arXiv.1702.00786}. These observations may reveal electromagnetic precursors associated with jet activity before merger~\citep{10.1103/PhysRevD.96.023004}.

The gravitational wave strain from massive black hole mergers at cosmological distances is:
\begin{equation}
h \sim \frac{G M_{\text{chirp}}}{c^2 d} \left(\frac{G M_{\text{chirp}} \pi f}{c^3}\right)^{2/3}
\end{equation}
where $M_{\text{chirp}} = (M_1 M_2)^{3/5}/(M_1 + M_2)^{1/5}$ is the chirp mass and $f$ is the gravitational wave~frequency.

\subsection{Technological and Methodological~Advances}
\unskip

\subsubsection{Computational~Challenges}
The study of black hole jets requires computational methods that can handle the extreme dynamic range and multi-physics nature of the problem~\citep{10.3847/0067-0049/225/2/22, 10.3847/1538-4365/ac9966}. Current simulations are limited by computational resources and numerical methods, preventing the study of jets across all relevant scales simultaneously~\citep{10.1186/s40668-017-0020-2, 10.1051/0004-6361/201935559}.

The computational challenge can be quantified by the range of scales:
\begin{equation}
\frac{L_{\max}}{L_{\min}} \sim \frac{10^6 r_g}{c/\omega_p} \sim 10^{12} \left(\frac{M_{\text{BH}}}{10^9 M_\odot}\right) \left(\frac{n}{10^6 \text{ cm}^{-3}}\right)^{1/2}
\end{equation}

Future advances in high-performance computing, including exascale systems and specialized architectures such as GPUs, will enable more detailed simulations of jet physics~\citep{10.3847/1538-4365/ab3922, 10.1029/2018JA025713, 10.1007/s11214-025-01142-0}. The~development of adaptive mesh refinement and multi-scale methods will allow simulations to capture both the small-scale microphysics and large-scale dynamics~\citep{10.3847/0067-0049/225/2/22}. Machine learning techniques may also play an increasingly important role in accelerating simulations and identifying patterns in complex datasets~\citep{10.1103/RevModPhys.91.045002, 10.48550/arXiv.1904.07248}.

The integration of different physical processes, including general relativity, magnetohydrodynamics, kinetic plasma physics, and~radiative transfer, remains a major challenge~\citep{10.3847/2041-8213/ab9532, 10.3847/0004-637X/829/1/11}. Hybrid methods that combine different approaches for different regions or scales offer a promising path forward~\citep{10.3847/1538-4365/aab114, 10.1103/PhysRevLett.118.055103}. The~development of standardized simulation codes and comparison projects will help validate these methods and identify their limitations~\citep{10.3847/1538-4365/ab29fd}.

\subsubsection{Data Analysis and~Interpretation}
The next generation of observational facilities will produce unprecedented amounts of data, requiring new methods for analysis and interpretation~\citep{10.3847/1538-4357/ab042c, 10.1088/1538-3873/aaecbe}. Machine learning techniques will be essential for identifying interesting events in large datasets and for extracting physical parameters from complex observations~\citep{10.48550/arXiv.1904.07248, 10.1086/668468}.

The data volume from LSST alone is estimated as:
\begin{equation}
V_{\text{data}} \sim 20 \text{ TB night}^{-1} \times 3650 \text{ nights} \sim 70 \text{ PB}
\end{equation}

The development of standardized data formats and analysis pipelines will facilitate comparison between different observations and theoretical models~\citep{10.1126/sciadv.aaz1310}. Virtual observatory tools will allow for researchers to access and analyze data from multiple facilities simultaneously~\citep{10.1093/mnras/stz3244}. The~integration of observational data with theoretical models through Bayesian inference and other statistical methods will impose more robust parameter estimation~\citep{10.1103/PhysRevLett.125.141104}.

The interpretation of multi-messenger observations requires sophisticated models that can predict the correlated signals across different messengers~\citep{10.1038/s42254-019-0101-z}. The~development of such models and the statistical methods needed to analyze multi-messenger data will be decisive for maximizing the scientific return of future observations~\citep{10.1088/978-0-7503-1369-8}.

\section{Conclusions}

Black hole jets remain one of the grand challenges of modern astrophysics, encompassing fundamental physics from quantum mechanics to general relativity and spanning scales from the event horizon to intergalactic distances. The~upcoming synergy of next-generation facilities—EHT, CTA, LSST, and~WEBT —together with theoretical advances in GRMHD, kinetic plasma simulations, and~radiative transfer, promises a transformative leap forward in our~understanding.

The multi-scale nature of jet physics requires interdisciplinary approaches that combine observational astronomy, theoretical physics, and~computational modeling~\citep{10.1146/annurev-astro-081817-051948}. The~questions outlined in this review span multiple areas of physics and astrophysics, from~the fundamental nature of spacetime near black holes to the large-scale structure and evolution of the universe. Addressing these questions will require unprecedented coordination between different observational facilities and theoretical~approaches.

The Event Horizon Telescope has already demonstrated the power of horizon-scale imaging in exploring jet launching mechanisms and black hole properties~\citep{10.3847/2041-8213/ab0ec7, 10.3847/2041-8213/abe71d}. Future developments, including multi-frequency observations, time-resolved imaging, and~space-based extensions, will supply even more detailed insights into the physics near the event horizon~\citep{10.48550/arXiv.1909.01411, 10.1126/sciadv.aaz1310}. The~combination of total intensity and polarization observations will reveal the magnetic field structure responsible for jet launching and~collimation.

The Cherenkov Telescope Array will revolutionize our understanding of particle acceleration in jets through unprecedented sensitivity to high-energy gamma rays~\citep{10.1142/10986}. Its ability to detect rapid variability and resolve extended emission will constrain acceleration mechanisms and energy dissipation processes. The~synergy with neutrino detectors may finally resolve the question of hadronic versus leptonic acceleration in jets~\citep{10.1093/mnras/stac3190, 10.1093/mnras/sty1852}.

The Vera C. Rubin Observatory will unlock statistical insights into jet duty cycles, triggering mechanisms, and~feedback efficiency through long-term monitoring of millions of AGN~\citep{10.48550/arXiv.0912.0201, 10.3847/1538-4357/ab042c}. Its discovery of rare transients will expand our understanding of jet formation beyond the traditional AGN paradigm. The~correlation of optical variability with multi-wavelength observations will reveal the connection between accretion and jet~activity.

The WEBT has moved blazar studies from phenomenology to quantitative astrophysics. It demonstrates that optical and multifrequency monitoring is not merely light curve collection, but~a powerful tool for ``remote sensing'' of relativistic plasma. Webb~\citep{10.3390/galaxies4030015} turbulent cell model provided the theoretical language to interpret the small-scale structures. The~2020 WEBT campaign on BL Lacertae provided the perfect, data-rich ``laboratory'' to apply this model at the micro-scale Webb \& Sanz~\citep{10.3390/galaxies11060108} and, simultaneously, to~discover how those turbulent structures coexist with and modify large-scale magnetohydrodynamic instabilities~\citep{10.1038/s41586-022-05038-9}. The~WEBT collaboration will provide the definitive evidence for turbulence and instability as the twin engines of blazar~variability.

On the theoretical front, the~development of more sophisticated numerical models that can bridge the gap between horizon-scale physics and large-scale propagation remains a key challenge~\citep{10.3847/0067-0049/225/2/22, 10.3847/1538-4365/ac9966}. The~incorporation of kinetic plasma physics into GRMHD simulations will be necessary for connecting theory with observations~\citep{10.1088/0004-637X/771/1/54, 10.1103/PhysRevLett.118.055103}. Advances in computational techniques and high-performance computing will permit more realistic simulations that can capture the full complexity of jet~physics.

The study of black hole jets also exemplifies the growing importance of multi-messenger astronomy~\citep{10.1038/s42254-019-0101-z}. The~combination of electromagnetic observations, gravitational waves, and~neutrinos represents complementary probes of jet physics that can break degeneracies inherent in single-messenger observations. Future detections of gravitational waves from black hole mergers may reveal new pathways for jet formation and illuminate insights into the role of magnetic fields in extreme gravitational~environments.

By connecting microphysics near the event horizon with the macroscopic impact of feedback on galaxies and clusters, the~study of black hole jets stands poised to illuminate fundamental processes of energy transport and cosmic evolution. The~next decade will likely see major breakthroughs in our understanding of these remarkable phenomena, with~implications extending far beyond the study of black holes~themselves.

The success of this endeavor will require continued investment in observational facilities, theoretical research, and~computational infrastructure~\citep{10.1146/annurev-astro-081817-051948}. It will also require fostering collaboration between different communities and disciplines, from~high-energy astrophysics to galaxy formation and cosmology. The~study of black hole jets exemplifies the interconnected nature of modern astrophysics and the power of combining multiple observational and theoretical approaches to address fundamental questions about the~universe.

As we stand on the threshold of a new era in jet astrophysics, the~convergence of revolutionary observational capabilities and sophisticated theoretical models promises to finally unlock the secrets of these cosmic accelerators. The~journey from the event horizon to the cosmic web, mediated by relativistic jets, represents one of nature's most extraordinary phenomena and continues to challenge our understanding of the fundamental laws of~physics.


\vspace{6pt} 





\authorcontributions{A.L.B.R. was specifically responsible for the Conceptualization, Supervision, funding acquisition and writing---original draft preparation of the work. N.M.N.d.R. undertook the tasks of Project
administration, writing---review and editing, and~Visualization. All authors have read and agreed to the published version of the~manuscript.}

\funding{\hl{This research was funded by CNPq (Conselho Nacional de Desenvolvimento Científico e Tecnológico), grant number 316317/2021-7.} 
} 

\dataavailability{\hl{No new data were created or analyzed in this study. Data sharing is not applicable to this article.}} 

\acknowledgments{\hl{The authors thank} 
 the referees for very helpful suggestions. N.M.N.d.R. thanks the support of PROBOL-UESC. 
\hl{A.L.B.R. also acknowledges the support from the CNPq (Conselho Nacional de Desenvolvimento Científico e Tecnológico) under Grant No. 404160/2025-5 (CNPq/MCTI Call No. 44/2024---Universal).}} 

\conflictsofinterest{The authors declare no conflicts of~interest.} 



\abbreviations{Abbreviations}{
The following abbreviations are used in this manuscript:\vspace{-6pt}

\noindent 
\begin{longtable}[l]{@{}ll} 
EHT & Event Horizon Telescope\\
CTA & Cherenkov Telescope Array\\
LSST & Vera C. Rubin Observatory's Legacy Survey of Space and Time\\
MADs & Magnetically Arrested Disks\\
GRMHD & General Relativistic Magnetohydrodynamic\\
AGN & Active Galactic Nuclei\\
H.E.S.S. & High Energy Stereoscopic System\\
MHD & Magnetohydrodynamic\\
PIC & particle-in-cell\\
TDEs & Tidal Disruption Events\\
LISA & Laser Interferometer Space Antenna\\
GRBs & Gamma-ray Bursts\\
SSC & Synchrotron Self-Compton\\
EC & External Compton\\
GPU & Graphics Processing Unit\\
UHECRs & Ultra-High-Energy Cosmic Rays\\
IceCube & Cubic-Kilometer Cherenkov Particle Detector\\
KM3NeT & Cubic Kilometre Neutrino Telescope\\
LIGO & Laser Interferometer Gravitational-Wave Observatory\\
WEBT & Whole Earth Blazar Telescope\\
JCMT & James Clerk Maxwell Telescope \\
CARMA & Combined Array for Research in Millimeter-wave Astronomy \\
SMT & Heinrich Hertz Submillimeter Telescope \\
SMA & Submillimeter Array \\
CSO & Caltech Submillimeter Observatory \\
APEX & Atacama Pathfinder Experiment \\
LMT & Large Millimeter Telescope \\
IRAM & Institute for Radio Astronomy in the Millimetre Range \\
SPT & South Pole Telescope\\

\end{longtable}
}


\begin{adjustwidth}{-\extralength}{0cm}

\reftitle{References}

\PublishersNote{}
\end{adjustwidth}
\end{document}